\documentclass[]{aa}  

\usepackage{graphicx}
\usepackage{xcolor}
\usepackage{float}
\usepackage{rotating}
\usepackage{booktabs}
\usepackage{mathtools}
\usepackage{placeins}
\usepackage{bm}
\usepackage{longtable}
\usepackage{adjustbox}
\usepackage{supertabular}
\usepackage{multirow}
\usepackage{tabularx}
\usepackage{url}
\usepackage{esdiff}
\usepackage{amsmath}
\usepackage{siunitx}
\usepackage[switch]{lineno}

\usepackage{hyperref}
\hypersetup{
  colorlinks=true,
    linkcolor=blue,
    citecolor=blue,
    urlcolor=magenta}

\usepackage{natbib}
\usepackage{txfonts}

\begin{document} 

   \title{Rotation of young solar-type stars as seen by \textit{Gaia} and K2}
    
   \titlerunning{Rotation of solar-type stars: a K2-\textit{Gaia} cross view}

   \author{S.N.~Breton\inst{1}
          \and 
          E.~Distefano\inst{1}
          \and
          A.C.~Lanzafame\inst{1,2}
          \and 
          D.B.~Palakkatharappil\inst{3}
          }
    \institute{INAF – Osservatorio Astrofisico di Catania, Via S. Sofia, 78, 95123 Catania, Italy \\
    \email{sylvain.breton@inaf.it}
    \and
    University of Catania, Astrophysics Section, Dept. of Physics and Astronomy, Via S. Sofia, 78, 95123, Catania, Italy
    \and
    Universit\'e Paris-Saclay, Universit\'e Paris Cit\'e, CEA, CNRS, AIM, 91191, Gif-sur-Yvette, France
    }

   \date{}

 \abstract
 {
 Accurate surface rotation measurements are crucial to estimate stellar ages and improve our understanding of stellar rotational evolution.
 Comparisons of datasets obtained from different space missions on common targets represent in this sense a way to explore the respective biases and reliability of the considered instruments, as well as a possibility to perform a more in-depth investigation of the properties of the observed stars.
 }
 {
 In this perspective, we aim at using observations for the K2 mission to provide an external validation to \textit{Gaia} rotation measurements, {\color{black} and confront observables available from \textit{Gaia}, K2, and \textit{Kepler}}.
 }
 {
 We therefore crossmatch the \textit{Gaia} rotation catalogue and the K2 mission Ecliptic Plane Input Catalogue (EPIC) in order to find \textit{Gaia} stars with both measured rotation and periods and available K2 light curves.
 Using our crossmatch, we analyse 1063 light curves from the K2 mission in order to characterise stellar rotational modulations and compare the recovered periods with \textit{Gaia} reference values. The K2/\textit{Gaia} cross-validated sample is used as a random-forest classifier training set to identify a subsample of \textit{Gaia} stars with similar properties.
 }
 {
 We validate the \textit{Gaia} rotation measurements for a large fraction of the sample and we discuss the possible origin of the discrepancies between some K2 and \textit{Gaia} measurements.
 We note that the K2 sample does not include members of the low-activity ultra-fast-rotating (UFR) population that was highlighted by \textit{Gaia} observations, a feature that we explain considering the instrumental capabilities of K2.
 Placing our sample in perspective with the full \textit{Gaia} rotation catalogues and \textit{Kepler} observations, we show that the population for which both \textit{Gaia} and K2 are able to measure rotation is composed of young late-type stars, a significant fraction of which is not yet converged on the slow-rotator gyrochronological sequence.
 {\color{black}In order to identify additional targets that have properties similar to the cross-validated K2 sample (considering in particular rotation and activity index), we compute the Local Outlier Factor (LOF) of the stars in the \textit{Gaia} DR3 rotation catalogue, considering the K2 stars as reference, and we identify 40,423 stars with a high degree of similarity, which can be useful for future statistical studies.}

 }
 { To the purpose of characterising the properties of young solar-type fast rotators, future photometric spaceborne missions such as PLATO will greatly benefit from the synergies with \textit{Gaia} observations that we illustrate in this work. 
 }


   \maketitle

\section{Introduction \label{section:introduction}}

Rotation plays a crucial role in the complex magnetohydrodynamical interplay responsible for the existence of a variety of dynamo regimes in the convective envelopes of solar-type and low-mass stars \citep[e.g.][]{Brun2017b}.
An important feature of this stellar envelope magnetism is the slow but continuous loss to the surrounding medium of angular momentum carried away by particle injected in the stellar wind, resulting in a secular braking of stellar rotation \citep{Skumanich1972}.
As stars spin down, they enter the so-called gyrochronological slow-rotator sequence \citep{Barnes2003,Barnes2007,Lanzafame2015} which allows a reliable estimation of their age, at least until they reach a transition regime where the efficiency of angular momentum transfer towards the stellar wind seems to weaken \citep[e.g.][]{vanSaders2016,Hall2021}.
Collecting accurate measurements of surface rotation from pre-main sequence to the beginning of the subgiant phase is therefore crucial to estimate stellar ages, to calibrate angular momentum loss rate at different evolutionary stage and to elucidate the properties of the magnetic mechanisms driving this phenomenon \citep[e.g.][]{Metcalfe2022}.   

Since its launch in 2013, the \textit{Gaia} mission \citep{Gaia2016} has provided surface rotation measurements for a vast sample of late-type stars \citep{Lanzafame2018,Distefano2023}, especially for young fast rotators, unravelling unexpected regimes of activity and transitions for ultra-fast rotators \citep[UFR,][]{Lanzafame2019}. Following the loss of the \textit{Kepler} satellite \citep{Borucki2010} second reaction wheel and the forced abandon of the initial \textit{Kepler} field in 2013, the K2 mission \citep{Howell2014} was designed in order to monitor the variability of stars located in pointing fields along the ecliptic plane.
In particular, the observations collected by K2 demonstrated that some features characterised by \textit{Kepler} for the rotational distribution of late-type stars such as the bimodality of rotation periods distribution \citep{McQuillan2014}, were not specific to this field of view \citep{Reinhold2020,Gordon2021}, supporting the hypothesis that the detected gap is connected with core-envelope coupling \citep{Spada2020,Lu2022}. The properties of the \textit{Kepler} rotational sample was extensively reviewed and put in perspective with other surveys by \citet{Santos2024}.

While the intersection between \textit{Kepler} stars and \textit{Gaia} targets with measured rotation is empty, some of the stars included in the \textit{Gaia} rotation  catalogue have been observed by the K2 mission.
Considering stars with both \textit{Gaia} and K2 time series allows an independent validation of stellar surface rotation measurements. 
In particular, it provides the opportunity to investigate the physical properties of the common targets and to compare the behaviour of the activity indexes obtained from each mission.
As \textit{Kepler} and K2 observations are very similar, this analysis also allows us to set in perspective the respective regions of the parameter space that \textit{Kepler}/K2 and \textit{Gaia} allowed unraveling.
Taking advantages of \textit{Gaia} multi-band photometry and infrared spectroscopy, it also opens the perspective to shed a multi-wavelength light on the otherwise mono-band \textit{Kepler}/K2 observations.  
We emphasise that, even if K2 probed an ensemble of fields with limited area when comparing it with the extent of the Transiting Exoplanet Survey Satellite \citep[TESS,][]{Ricker2015}, the one-meter telescope aboard the \textit{Kepler} satellite enables observations of fainter stars than for TESS. 

In this paper, we study the sample of \textit{Gaia} solar-type rotators for which K2 light curves are available. In particular, we explore how combining the multi-band \textit{Gaia} observations with the quasi-continuous high precision photometric data from K2 allows us to constrain their rotation to activity relation. 
The layout of the paper is as follows.
In Sect.~\ref{sec:data_analysis}, we present our method to crossmatch entries between \textit{Gaia} DR2, DR3 and K2 targets, before detailing the analysis we apply on K2 recovered light curves to measure stellar surface rotation. 
We show how the K2 sample compare with the full \textit{Gaia} DR3 rotation catalogue and the \textit{Kepler} sample of rotators in terms of activity indexes and other parameters collected through \textit{Gaia} observations.
In Sect.~\ref{sec:results}, we combine \textit{Gaia} and K2 observations to explore the properties of the stars for which we were able to measure a reliable rotation period from the K2 light curve. 
In Sect.~\ref{sec:discussion}, we discuss the discrepancies observed between \textit{Gaia} and K2 rotation measurements and we present a machine-learning-based analysis to attribute a reliability score to the \textit{Gaia} rotators similar to the stars from the K2 reference sample.
In Sect.~\ref{sec:conclusion} we present the conclusions and perspective of our analysis.

\section{Data analysis \label{sec:data_analysis}}

\subsection{\textit{Gaia}/K2 crossmatch}

\begin{figure*}[ht!]
    \centering
    \includegraphics[width=\textwidth]{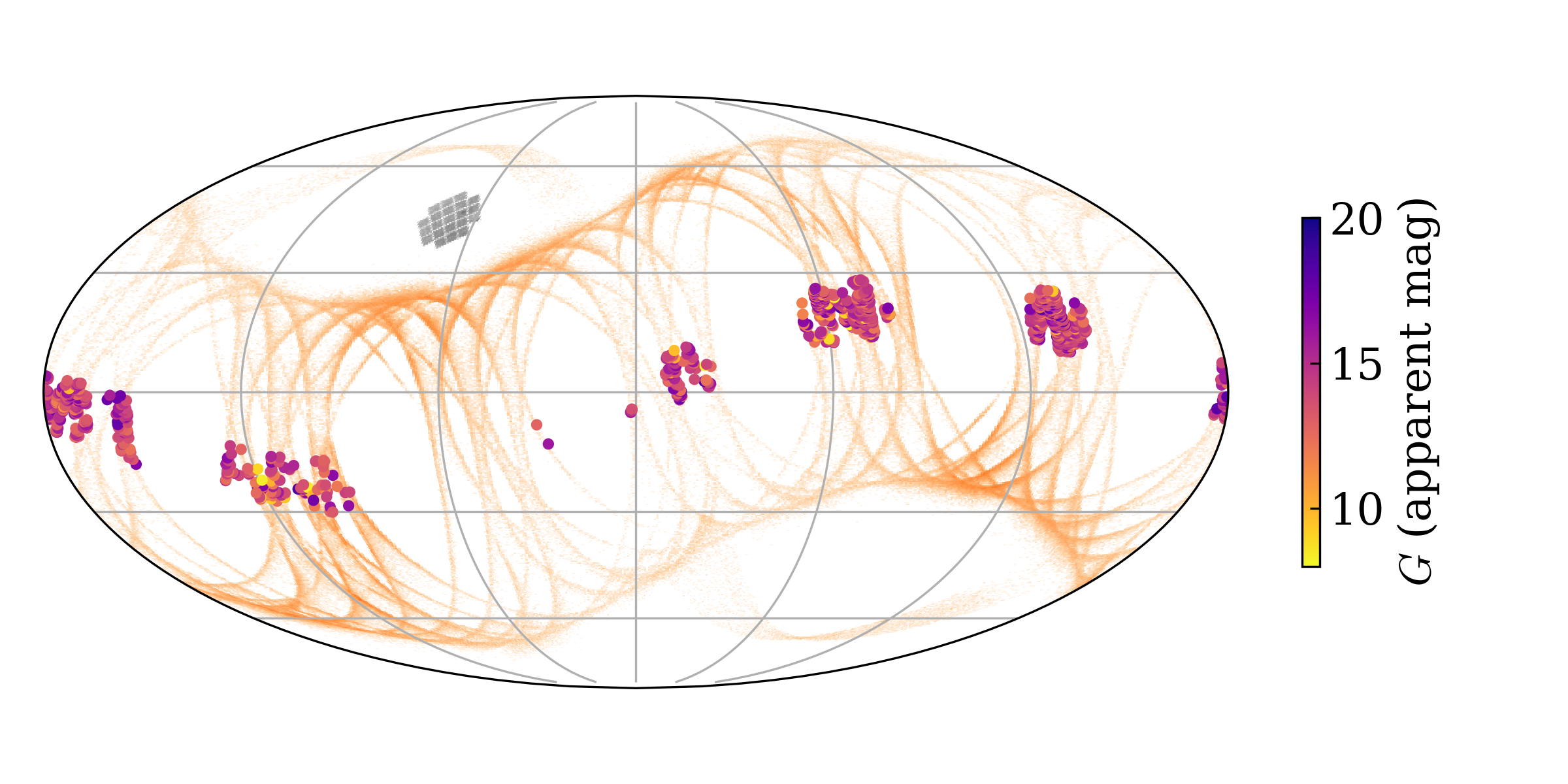}
figur    \includegraphics[width=\textwidth]{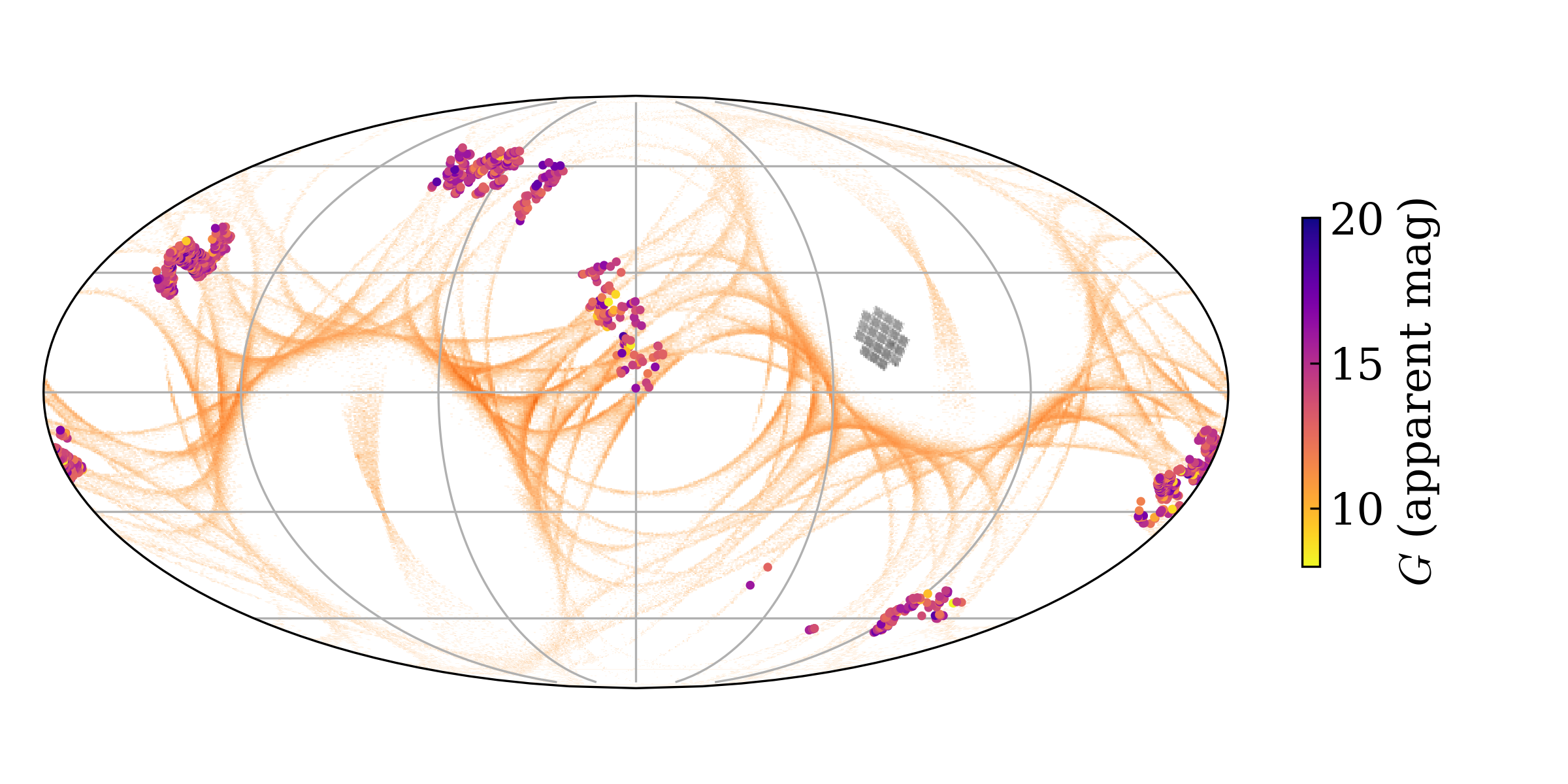}
    \caption{\textit{Top}: Right ascension and declination map distribution of the \textit{Gaia} targets possessing both a \textit{Gaia} rotation period from DR2/DR3 and an EVEREST K2 light curve. The magnitude $G$ is colour-coded. The full sample of \textit{Gaia} targets with a rotation period in DR3 is shown for reference \citep{Distefano2023} with their density colour-coded in orange. The location of the \textit{Kepler} field is shown in grey.
    \textit{Bottom:} Same for the galactic location of the targets.
    }
    \label{fig:galactic_location}
\end{figure*}

Given that some stars with detected rotational modulation from \textit{Gaia} DR2 \citep{Gaia2018} are missing in \textit{Gaia} DR3 \citep{GaiaDR3}, it is necessary to consider rotation catalogues for both data releases in order to obtain an extensive crossmatch with K2 observations.

We started by considering the crossmatch catalogue between K2 Ecliptic Plane Input Catalog (EPIC) identifiers and \textit{Gaia}~DR2 identifiers \footnote{The K2-\textit{Gaia} DR2 crossmatch is available here: \url{https://gaia-kepler.fun}.}. We then collected the \textit{Gaia} DR3 identifiers of the targets using the DR2 neighbourhood table from \textit{Gaia} Early Data Release 3 \citep[EDR3,][]{GaiaEDR3_2021}, considering each time the closest neighbour found for each \textit{Gaia} DR2 star.
By crossmatching this list with the DR2 \citep{Lanzafame2018} and DR3 \citep{Distefano2023} rotation catalogues\footnote{\color{black}See the online documentation of the \texttt{vari\_rotation\_modulation} table: \url{https://gea.esac.esa.int/archive/documentation/GDR3//Data_analysis/chap_cu7var/sec_cu7var_rotmod}.} we recovered 1080 K2 stars with a \textit{Gaia} rotation measurement.
We gave priority to the DR3 rotation period when available. 
We underline that we were able to assign to each K2 target a unique \textit{Gaia} DR2 and DR3 identifier.
Some of the targets are member of known open cluster (see Appendix~\ref{sec:open_clusters}).

To carry out the subsequent analysis, we chose to consider the EPIC Variability Extraction and Removal for Exoplanet Science Targets \citep[EVEREST,][]{Luger2016} light curves, which are publicly available. 
The EVEREST data reduction process was designed for a versatile range of science case, among which the analysis of stellar variability.
We recovered the EVEREST light curves from the Mikulski Archive for Space Telescopes (MAST) using the interface provided by the \texttt{lightkurve} module \citep{Lightkurve2018}.
We were able to download EVEREST light curves  for 1063 targets among the 1080 stars in our crossmatch.
We show in Fig.~\ref{fig:galactic_location} the sky location of the stars with recovered light curves, as well as their galactic locations. 
We compare this distribution to the population of stars with \textit{Gaia} DR3 rotation periods.
The distribution of K2 targets along the ecliptic plane and successive K2 pointings is clearly visible. 
For comparison, we also show the location of the \textit{Kepler} field. As already mentioned, the field is outside of the sky region where the \textit{Gaia} photometric observations allow a measurement of rotation.

\subsection{Rotation analysis for K2 light curves \label{sec:rotation_analysis}}

\begin{figure}[ht!]
    \centering
    \includegraphics[width=0.49\textwidth]{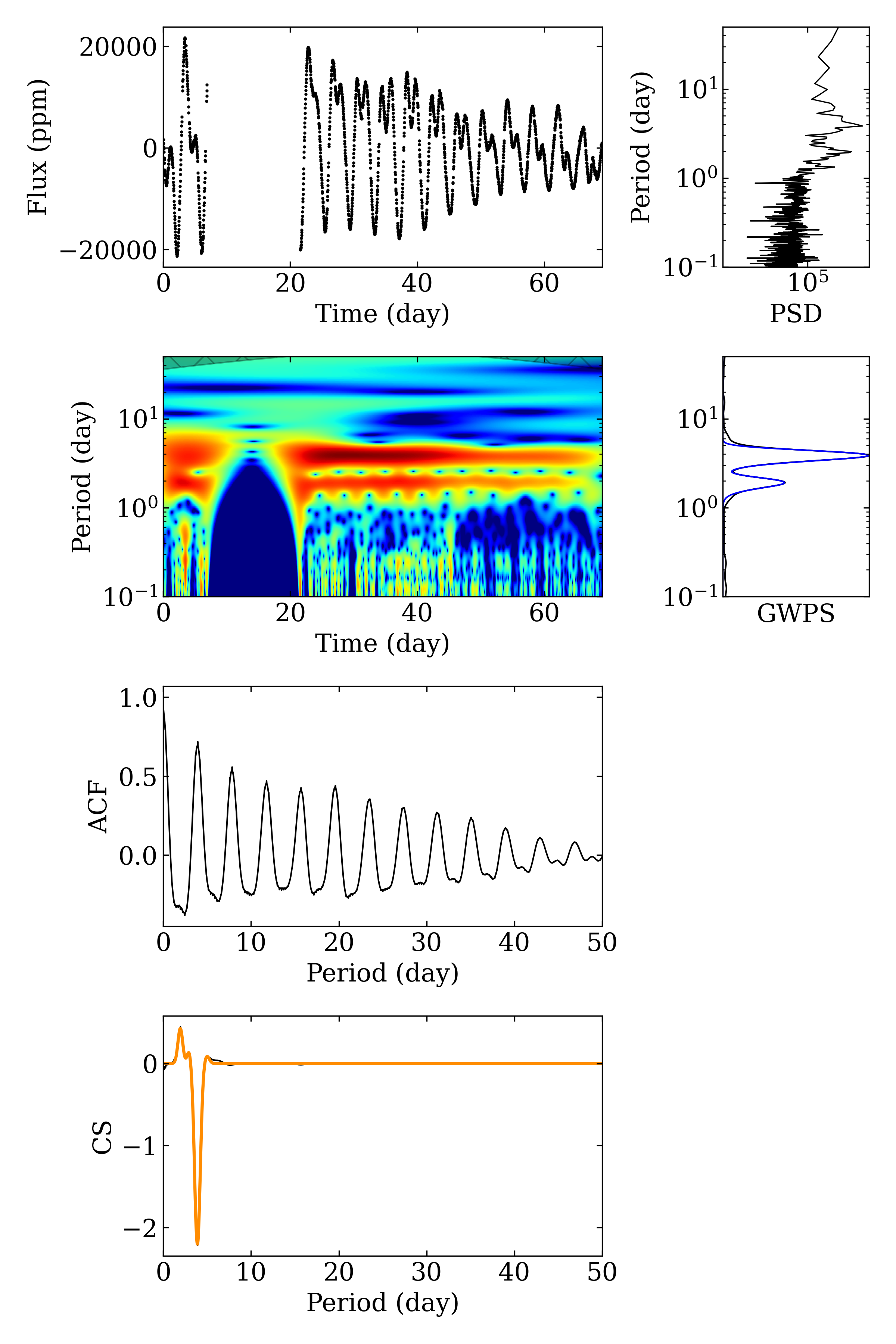}
    \caption{\textit{From top to bottom:} Light curve (\textit{left}) and power spectral density (\textit{right}), WPS (\textit{left}) and GWPS (\textit{right}), ACF, and CS for EPIC~201121691. The set of Gaussian profiles fitted on the GWPS and the CS are shown in blue and orange, respectively.}
    \label{fig:example_rotation_analysis}
\end{figure}

In order to apply our rotation measurement methodology to the EVEREST light curves, we discard the data points with bad quality flags. In order to keep an evenly sampled time series, these data points are replaced by zero-values. We also have to deal with stars that were observed in several K2 sectors. In this case, before renormalising the light curves to zero-mean part-per-million time series, we adjust the mean level of each segment so it is similar in each sector of observations. 

The procedure to measure rotation periods in the K2 light curves is similar to the one described in \citet{Santos2019,Santos2021} and combines several analysis methods in order to ensure robustness of the recovered periods \citep{Aigrain2015}. 
The analysis was performed with the Python module \texttt{star-privateer}\footnote{The source code is accessible at \url{https://gitlab.com/sybreton/star_privateer} and the corresponding documentation is hosted at \url{https://star-privateer.readthedocs.io/en/latest}.} \citep{Breton2024PLATO}, originally developed as a demonstrator module for the stellar rotation and activity algorithms to be implemented in the Planetary Transit and Oscillations of Stars \citep[PLATO,][]{Rauer2025} mission pipeline.

Global wavelet power spectrum (GWPS) of the sixth order Morlet wavelet decomposition \citep{Torrence1998,Liu2007,Mathur2010} of the time series is combined with the light curve auto-correlation function \citep[ACF,][]{McQuillan2013b} to compute the composite spectrum \citep[CS,][]{Ceillier2017}. A candidate rotation period is extracted from each of the three methods (GWPS, ACF, and CS). In the case of the GWPS and CS, the period is determined by fitting a set of Gaussian profile on the spectrum. The central period of the profile with largest amplitude is taken as the candidate period. In the case of the ACF, we consider the first local maximum, or the second one if it is at least twice higher than the first.
To mitigate edge effects in the wavelet transform, the time series was zero-padded before computing the wavelet power spectrum (WPS) and the GWPS. This multi-method analysis is illustrated in Fig.~\ref{fig:example_rotation_analysis} in the case of EPIC~201121691 for which the final recovered rotation period is 3.9 days, consistent with the 4.2 days period provided in the \textit{Gaia} catalogue.
A gap in the light curve is visible between 10 and 20 days. This is due to the fact that we set to zero measurements with bad quality flags, which occurs for a significant fraction of the light curve we analyse. Nevertheless, as visible in the panels displaying the WPS, GWPS, ACF, and CS, this does not bias the recovery of the rotation periods in the different methodology, as these methods are robust to the existence of missing data. Indeed, null values are not changing the ACF profile while the convolution process of the wavelet decomposition is only marginally affected by the gaps. Only it can happen that, when the convoluting wavelet is of similar timescale than the gap, the WPS exhibits locally a spurious feature.  

The final rotation period is selected by the Random Forest over Stellar Rotation methodology \citep[ROOSTER,][]{Breton2021}, together with the attribution of a rotation score related to the reliability of the rotational modulation detected in the light curve. 
This rotation score takes values between 0 and 1 and is computed as the fraction of decision trees is in the forest that attributes the detected-rotation label to the light curve.
The ROOSTER random forest classifiers were trained with data obtained from a subset of the \citet{Santos2019,Santos2021} \textit{Kepler} rotation catalogue. 
In order to adapt the training parameter space to the fact that we expect to recover an important population of very fast rotators, {\color{black}, we included in the training a significant fraction of fast rotating stars, and we subdivided all included light curves in chunks of 90 days to match the length of the K2 light curves.} 
The performances of the classifiers were then validated with an independent subset of \textit{Kepler} light curves. 
Considering this test set, the correct rotation period was recovered in 94.2\% of the cases, and, assuming a 0.5-rotation-score detection threshold, rotating stars were correctly detected with a 93.2\% accuracy.
We recall that the main source of error at this step of the analysis usually comes from {\color{black}confusion between the fundamental period and its first overtone}, residual modulations from instrumental features ({\color{black}mostly the quarterly modulation from \textit{Kepler}}), or light curve contamination by background or neighbourhood stars. 
{\color{black}
The \textit{Kepler} quarterly modulation manifests itself mainly at 40-50 days and is an issue mostly when it has to be distinguished from a low-amplitude Sun-like signal with weak coherence over time. As we expect to detect mostly active fast rotators in our sample, this should not be a critical issue for the analysis of the K2 light curves. 
Though it was found that some instrumental features differ from \textit{Kepler} to K2 \citep{Moreno2021}, we are able to show below (see Appendix~\ref{sec:comparison_with_rh}) that it does not affect the ROOSTER ability to correctly determine rotation periods. 
More details on the procedure we developed to adapt the ROOSTER training set for the analysis of K2 light curves can be found in Appendix~\ref{sec:rooster_training_set}.}  

Once the rotation period has been measured, it is possible to compute the $S_\mathrm{ph}$ photometric activity proxy \citep{Mathur2014,Mathur2014b} as the average of the standard deviations computed on time series segments with length $5 \times P_\mathrm{rot}$. 
The $S_\mathrm{ph}$ proxy was demonstrated to be strongly correlated with other standard activity indicators \citep{Salabert2017}. Combined with the rotation periods, it enables a reliable estimation of the stellar age through the so-called magneto-gyrochronology method \citep{Mathur2023}.

\section{Results \label{sec:results}}

\subsection{Rotation period in K2 vs Gaia \label{sec:rotation_period_k2_gaia}}

\begin{figure}[ht!]
    \centering
    \includegraphics[width=0.49\textwidth]{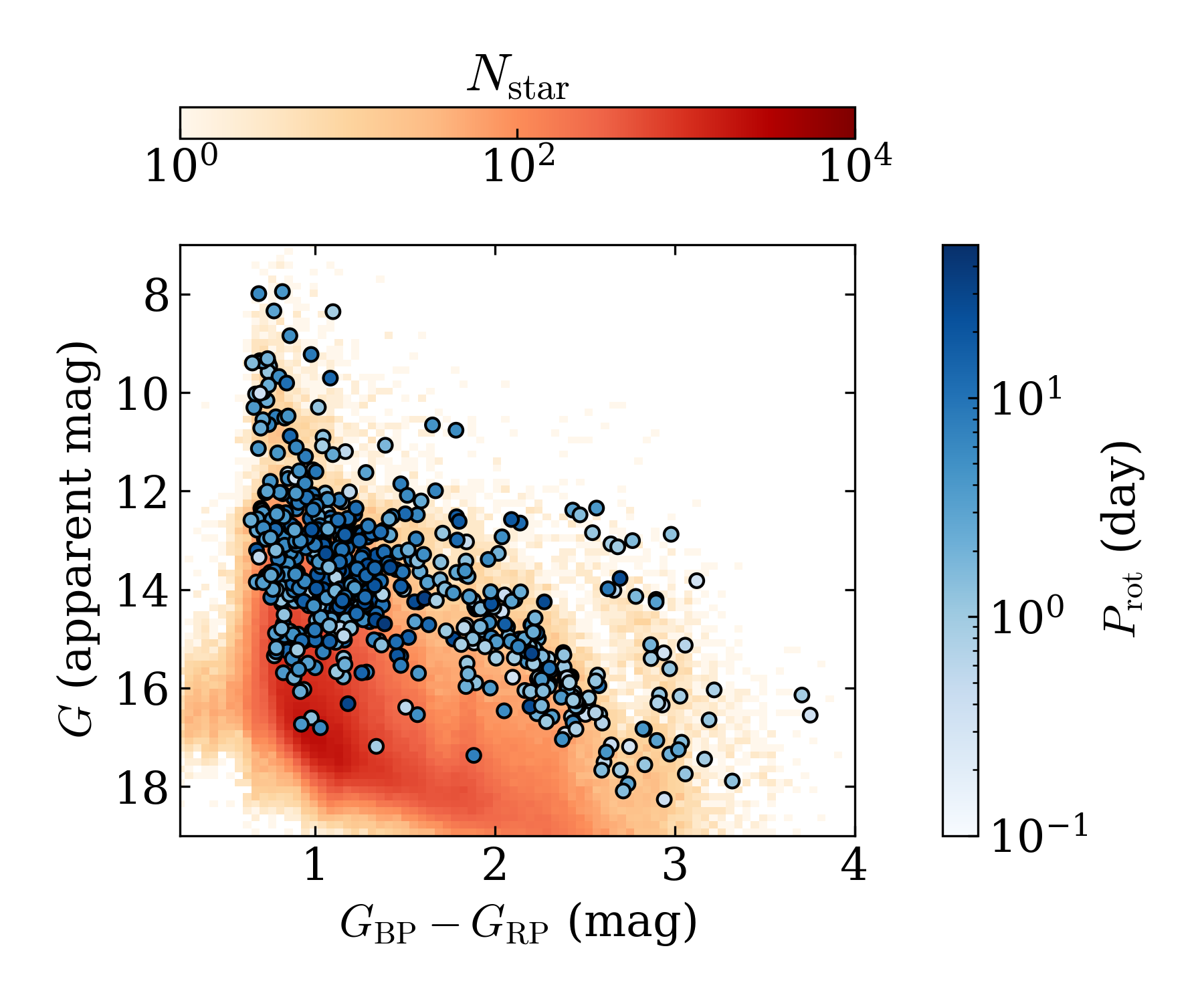}
    \includegraphics[width=0.49\textwidth]{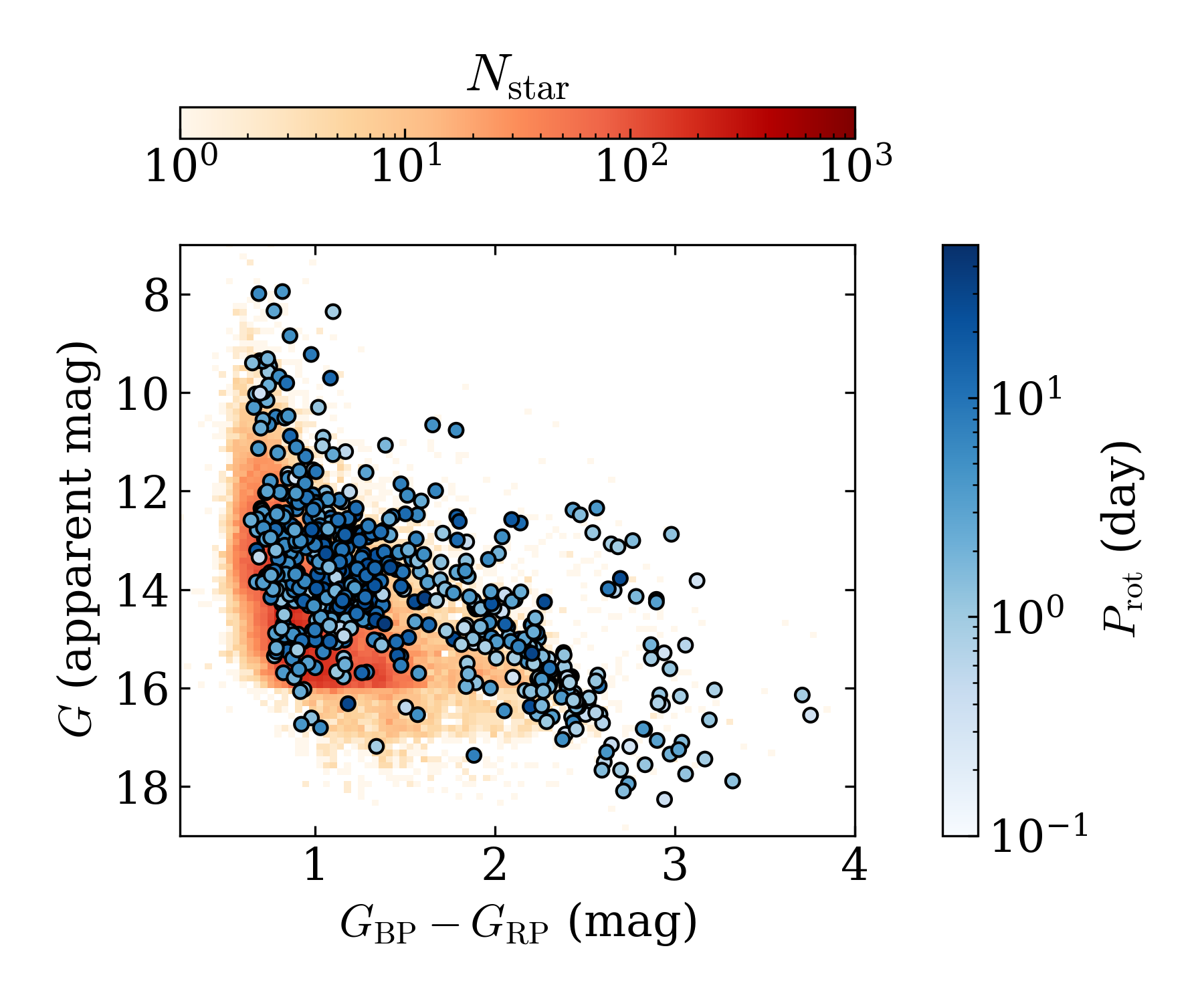}
    \caption{Apparent magnitude $G$ vs $G_{\rm BP} - G_{\rm RP}$ diagram. $P_{\rm rot}$ is colour-coded for the stars where the rotation period measured between K2 and \textit{Gaia} is cross-validated. In the top panel, the density distribution of stars with a \textit{Gaia} rotation period is shown, while on the bottom panel, our sample is compared to the distribution of the \textit{Kepler} stars for which \citet{Santos2019,Santos2021} provided a rotation period.}
    \label{fig:bp_rp_relative_gmag_diagram}
\end{figure}

\begin{figure}[ht!]
    \centering
    \includegraphics[width=0.49\textwidth]{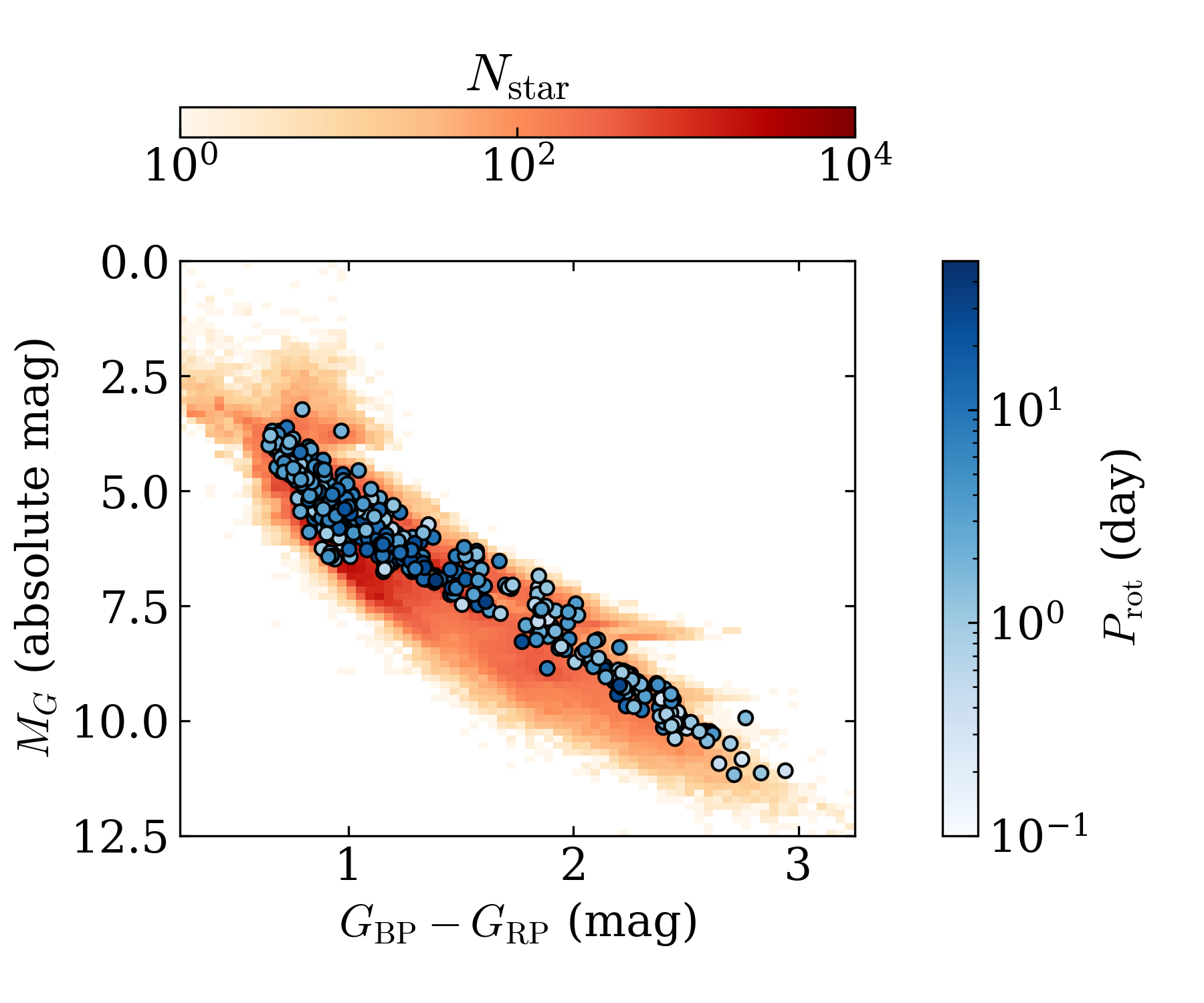}
    \includegraphics[width=0.49\textwidth]{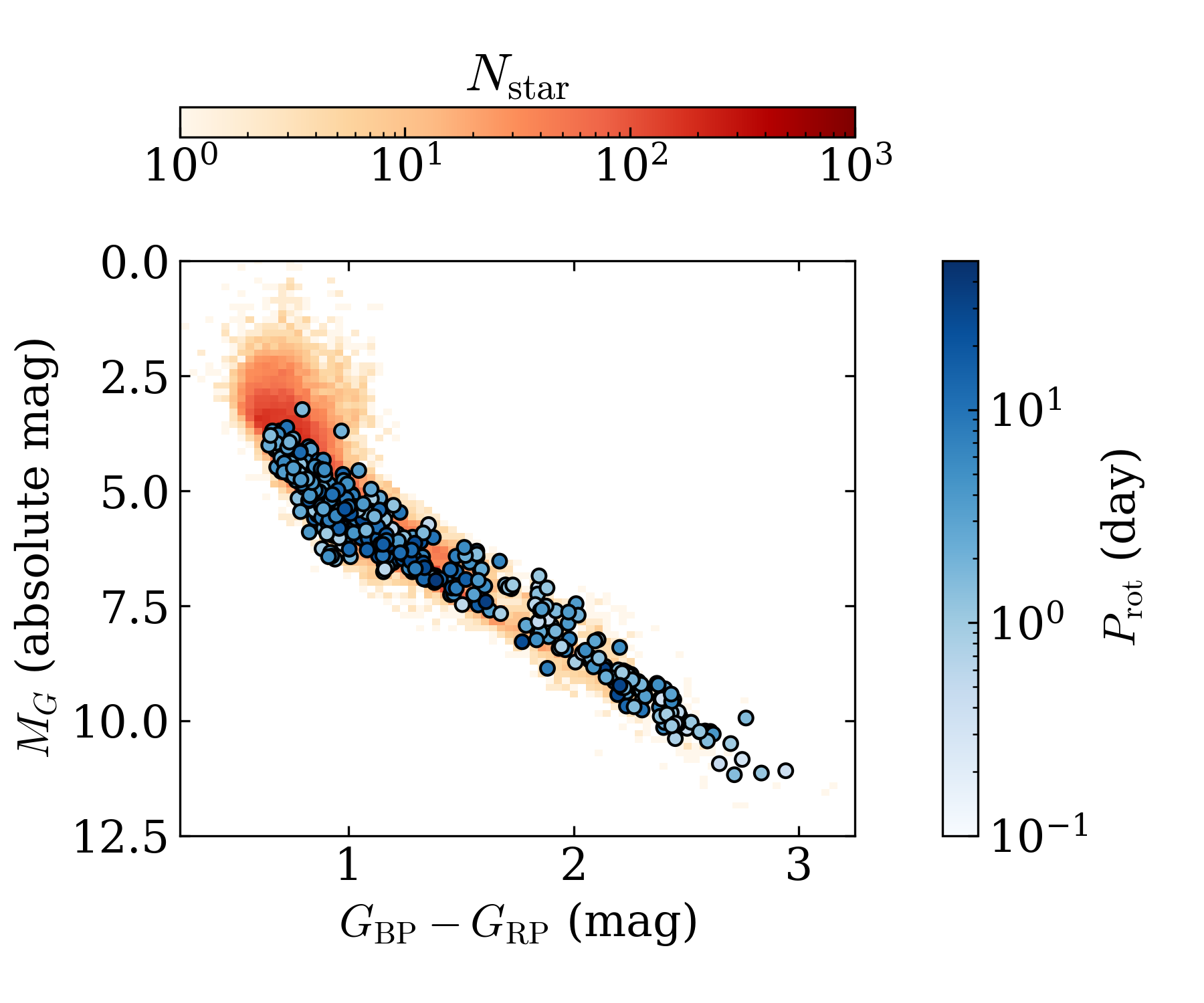}
    \caption{Absolute magnitude $M_G$ vs $G_{\rm BP} - G_{\rm RP}$ diagram. $P_{\rm rot}$ is colour-coded for the stars where the rotation period measured between K2 and \textit{Gaia} is cross-validated. In the top panel, the density distribution of stars with a \textit{Gaia} rotation period is shown, while on the bottom panel, our sample is compared to the distribution of the \textit{Kepler} stars for which \citet{Santos2019,Santos2021} provided a rotation period.}
    \label{fig:bp_rp_gmag_diagram}
\end{figure}

\begin{figure*}[ht!]
    \centering
    \includegraphics[width=\textwidth]{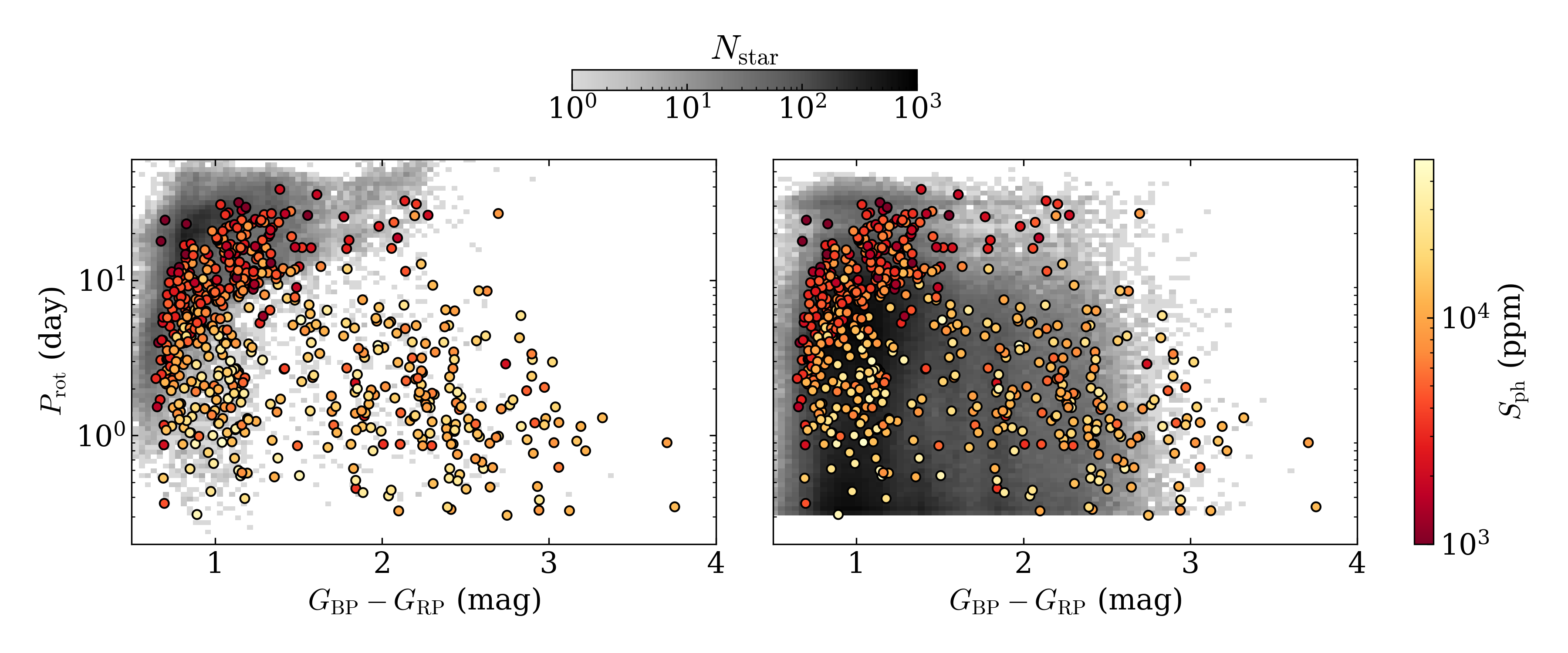}
    \caption{$P_\mathrm{rot}$ vs $G_{\rm BP} - G_{\rm RP}$ diagram. The $S_{\rm ph}$ value is colour-coded for the stars for which \textit{Gaia} and K2 rotation measurements are cross-validated. On the left panel, the sample is compared with the density distribution of the \textit{Kepler} stars from \citet[][in grey]{Santos2019,Santos2021} while on the right panel it is compared with the density distribution of the stars from \textit{Gaia} DR3 (in grey).}
    \label{fig:bp_rp_prot_diagram_combined.png}
\end{figure*}

Among the sample of 1063 stars we analysed, we are able to extract 598~periods laying between 0.45 and 2.1 times the \textit{Gaia} value and achieving a ROOSTER rotation score above 0.5.
This validation criterion is deliberately relaxed. 
In particular, it is aimed at accounting for the fact that most of the observations are not contemporaneous, meaning that the observable rotation periods may experience significant variations due to the latitudinal migration of active regions with time.
We will refer later to this sample as the cross-validated sample.
For 146 more stars, the K2 rotation period is outside of the range defined above but the ROOSTER rotation score is above 0.9, supporting a reliable detection of rotation. 
As there is no independent validation of the K2 measurement from the \textit{Gaia} observations for these stars, we require a larger ROOSTER score in order to keep these stars in our sample.
We will discuss further the properties of these stars in Sect.~\ref{sec:inconsistence_k2_gaia}.
Among this sample of 744 stars, 522 have reference \textit{Gaia} measurements from DR3 and we use values from \textit{Gaia} DR2 for the 222 others. 
Their properties are summarised in Table~\ref{table:summary_table}.
For the 319 remaining stars, visual inspection of their light curves and the results of the rotational analysis confirms that the quality of the K2 observations is not sufficient to extract any useful observables for these targets. We therefore do not consider these remaining stars in what follows. 
Also, unless mentioned otherwise, the $P_\mathrm{rot}$ we display for the stars in our sample is the one obtained from the K2 light curves.
{\color{black}In order to further validate the reliability of our K2 measurements, we compare our results with the ones obtained by \citet{Reinhold2020}. Among our 744 stars, they published a rotation measurement for 337 targets, for which we obtain a very good agreement, beyond 95\%. (see Appendix~\ref{sec:comparison_with_rh}).}

In Fig.~\ref{fig:bp_rp_relative_gmag_diagram}, we compare the location of the stars in this sample in the diagram $G_{\rm BP} - G_{\rm RP}$ vs apparent $G$~magnitude to the \textit{Gaia} DR3 stars with rotation periods and to the \textit{Kepler} sample from \citet{Santos2019,Santos2021}. 
The faint star cutoff is clearly visible \textit{Kepler} stars distribution as the distribution density abruptly decreases beyond $G=16$. This cutoff is also visible in our K2 sample, we note that most of the targets beyond it are M-type stars, consistently with the fact that \textit{Gaia} is able to probe a population of stars much more distant and faint than the bulk population included in the \textit{Kepler} Input Catalog (KIC) and the EPIC. 
We are using the $G_{\rm BP} - G_{\rm RP}$ measurements from \textit{Gaia} DR3, which are available for the entirety of our sample except four targets. As measurements from \textit{Gaia}~DR2 and DR3 are not homogeneous, these four stars are not displayed on the figures. 
In this figure and all that follow, we correct, when possible, the $G_{\rm BP} - G_{\rm RP}$ value by subtracting the dereddening coefficient computed by the \textit{Gaia} astrophysical parameter inference system \citep[Apsis,][]{Bailer-Jones2013,Creevey2023,Fouesneau2023}.

The magnitude cutoff also has an impacts when considering the absolute magnitude $M_G$ computed by the Apsis, as shown in Fig.~\ref{fig:bp_rp_gmag_diagram}. 
The $M_G$ distribution of our sample is very similar to the \textit{Kepler} one, except for the group of brighter stars visible on the upper-left corner: we remind that \citet{Santos2021} considered subgiant stars that are not considered in the \textit{Gaia} rotation analysis. 
We also note that, at a given $G_{\rm BP} - G_{\rm RP}$ colour, the $M_G$ range is wider for the \textit{Gaia} population than for the stars observed by \textit{Kepler}/K2, including significantly more stars that are less bright for a given $G_{\rm BP} - G_{\rm RP}$ colour.

In Fig.~\ref{fig:bp_rp_prot_diagram_combined.png}, we show the $G_{\rm BP} - G_{\rm RP}$ vs $P_\mathrm{rot}$ diagram for our sample, compared with the reference distributions of \textit{Kepler} and \textit{Gaia}~DR3 targets. This diagram is useful to characterise the rotational regime in which the K2-\textit{Gaia} crossmatch stars are located. 
The \textit{Gaia} survey is shifted towards short periods with comparison to the \textit{Kepler} measurements. 
The short-period cut-off of $\sim$0.3 days used for \textit{Gaia} is visible. We also note that the bi-modality of the \textit{Kepler} distribution \citep{McQuillan2014}, also detected in K2 \citep{Gordon2021} is not apparent in \textit{Gaia}, probably because the feature is blurred by the reduced precision of \textit{Gaia} measurements in this range of periods.
Most of the F- and G- type stars from the K2 sample occupy a region in the $G_{\rm BP} - G_{\rm RP}$ vs $P_\mathrm{rot}$ parameter space that corresponds to the young and fast rotators observed in \textit{Kepler} for these given spectral types.
This population corresponds to one of the two bulks of rotators characterised by \textit{Gaia} for these spectral types, the second one being composed of much faster rotators. 
Concerning latest-types, it is visible by comparing the K2-periods with the \textit{Kepler} distribution that most of the K- and M-type stars from the sample have not converged yet on the slow-rotator gyrochronological sequence. 
As expected, the $S_\mathrm{ph}$ values are strongly anti-correlated with $P_\mathrm{rot}$, increasing as $P_\mathrm{rot}$ decreases. 

\subsection{Photometric activity index comparison \label{sec:photometric_activity_comparison}}

\begin{figure}[ht!]
    \centering
    \includegraphics[width=0.49\textwidth]{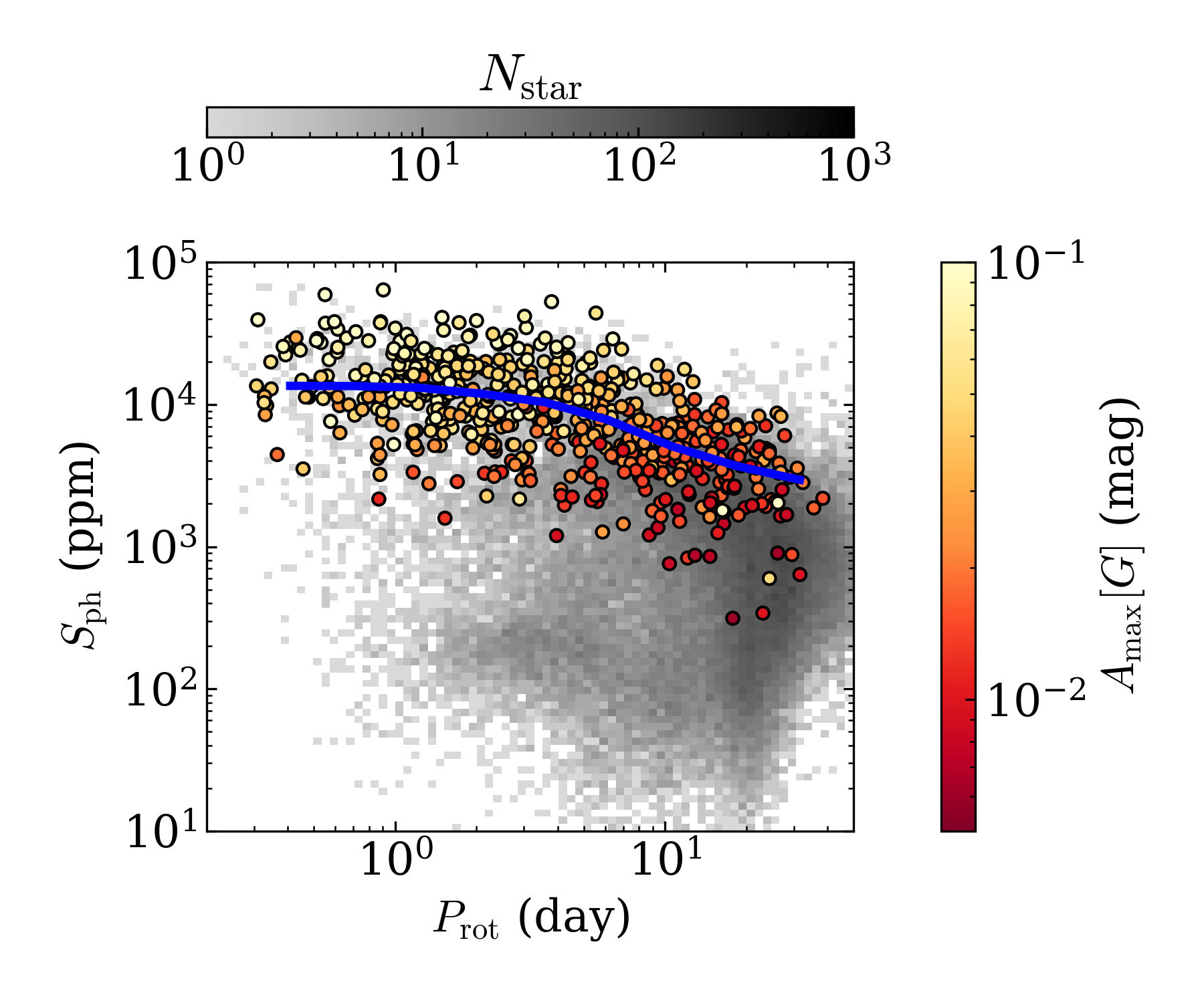}
    \caption{$S_\mathrm{ph}$ vs $P_{\rm rot}$ diagram for the stars where the rotation period measured between K2 and \textit{Gaia} is cross-validated. 
    The $Gaia$ activity index $A_{\rm max}[G]$ is colour-coded.
    The \textit{Kepler} distribution from \citet{Santos2019,Santos2021} is shown in grey for comparison.
    {\color{black} The median value of the binned distribution along $P_{\rm rot}$ is shown in blue.}
    }
    \label{fig:kepler_distribution}
\end{figure}

\begin{figure}[ht!]
    \centering
    \includegraphics[width=0.49\textwidth]{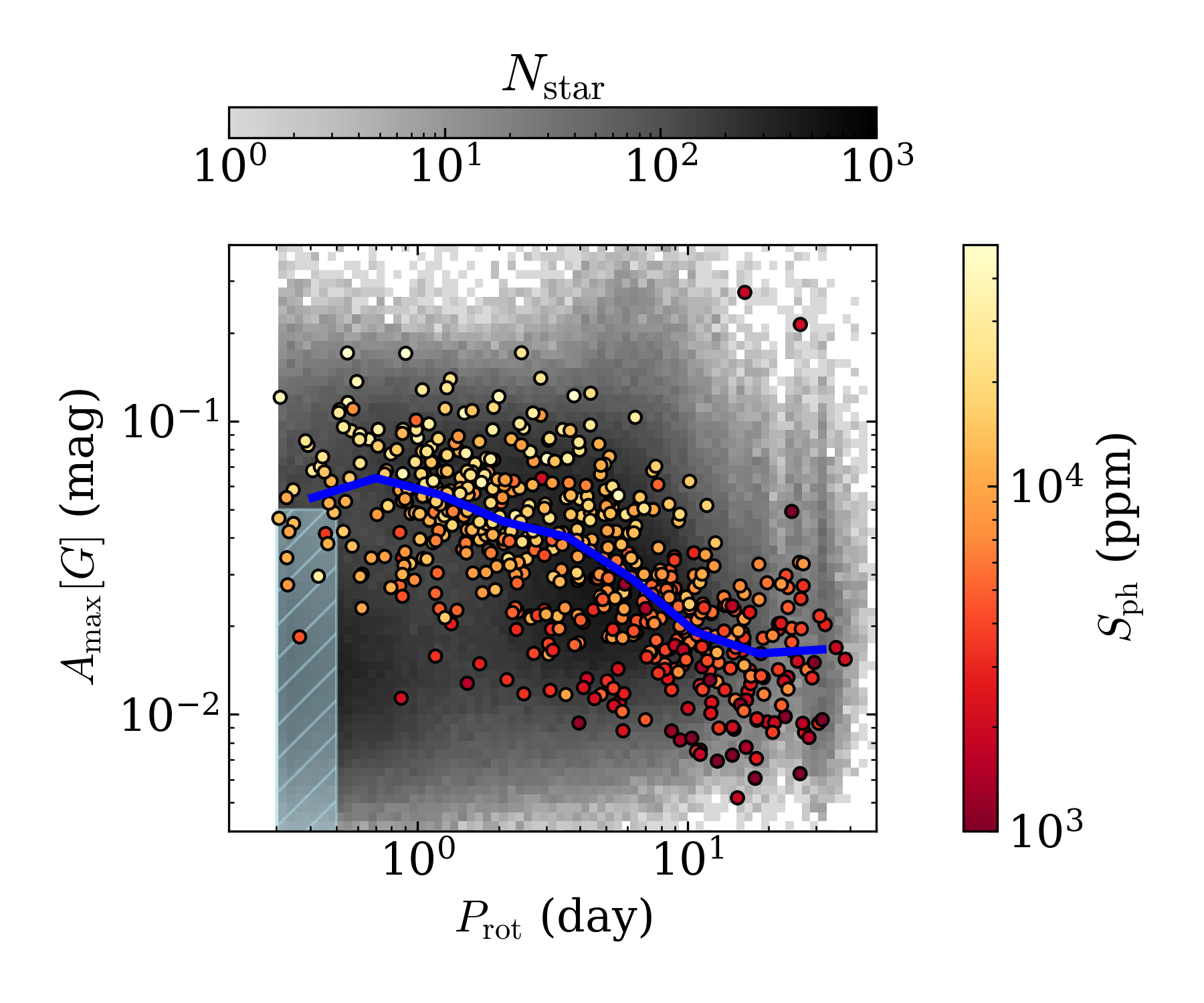}
    \caption{$A_{\rm max}[G]$ vs $P_{\rm rot}$  diagram for the stars where the rotation period measured between K2 and \textit{Gaia} is cross-validated. The $S_\mathrm{ph}$ is colour-coded.
    The full \textit{Gaia} DR3 distribution from \citet{Distefano2023} is shown in grey for comparison.
     {\color{black} The median value of the binned distribution along $P_{\rm rot}$ is shown in blue. The hatched blue area correspond to the UFR criterion selection from \citet{Lanzafame2019}.}
    }
    \label{fig:gaia_distribution}
\end{figure}

\begin{figure*}[ht!]
    \centering
    \includegraphics[width=\textwidth]{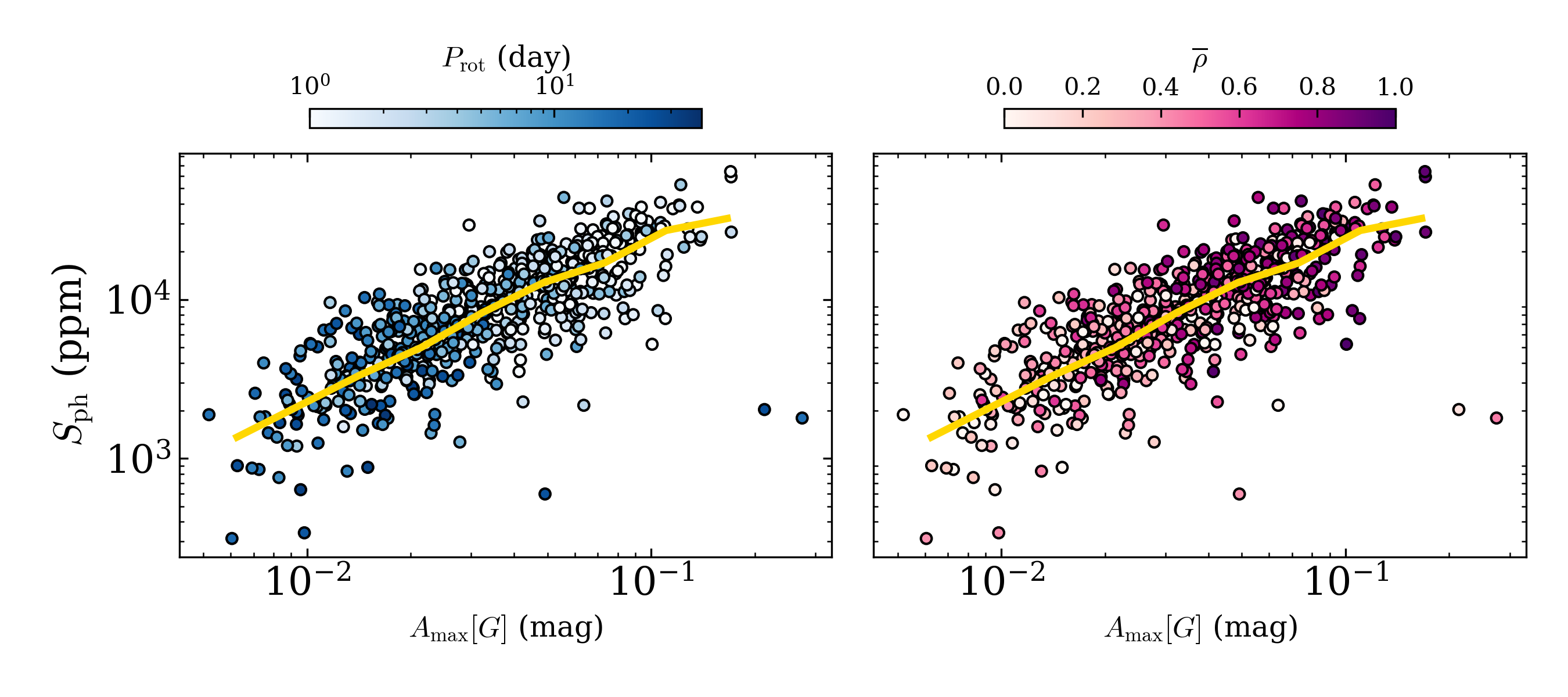}
    \caption{$S_\mathrm{ph}$ vs $A_{\rm max}[G]$ diagram for the stars where the rotation period measured between K2 and \textit{Gaia} is cross-validated. On the left panel $P_{\rm rot}$ are colour-coded while the mean {\color{black}Pearson} correlation coefficient $\overline{\rho}$ is colour-coded on the right panel.
    For readability, $\overline{\rho}$ values are colour-coded only on the 0 to 1 range. 
    {\color{black} On both panels, the median value of the binned distribution along $A_{\rm max}[G]$ is shown in blue.}
    }
    \label{fig:sph_vs_max_g}
\end{figure*}

\begin{figure*}[ht!]
    \centering
    \includegraphics[width=\textwidth]{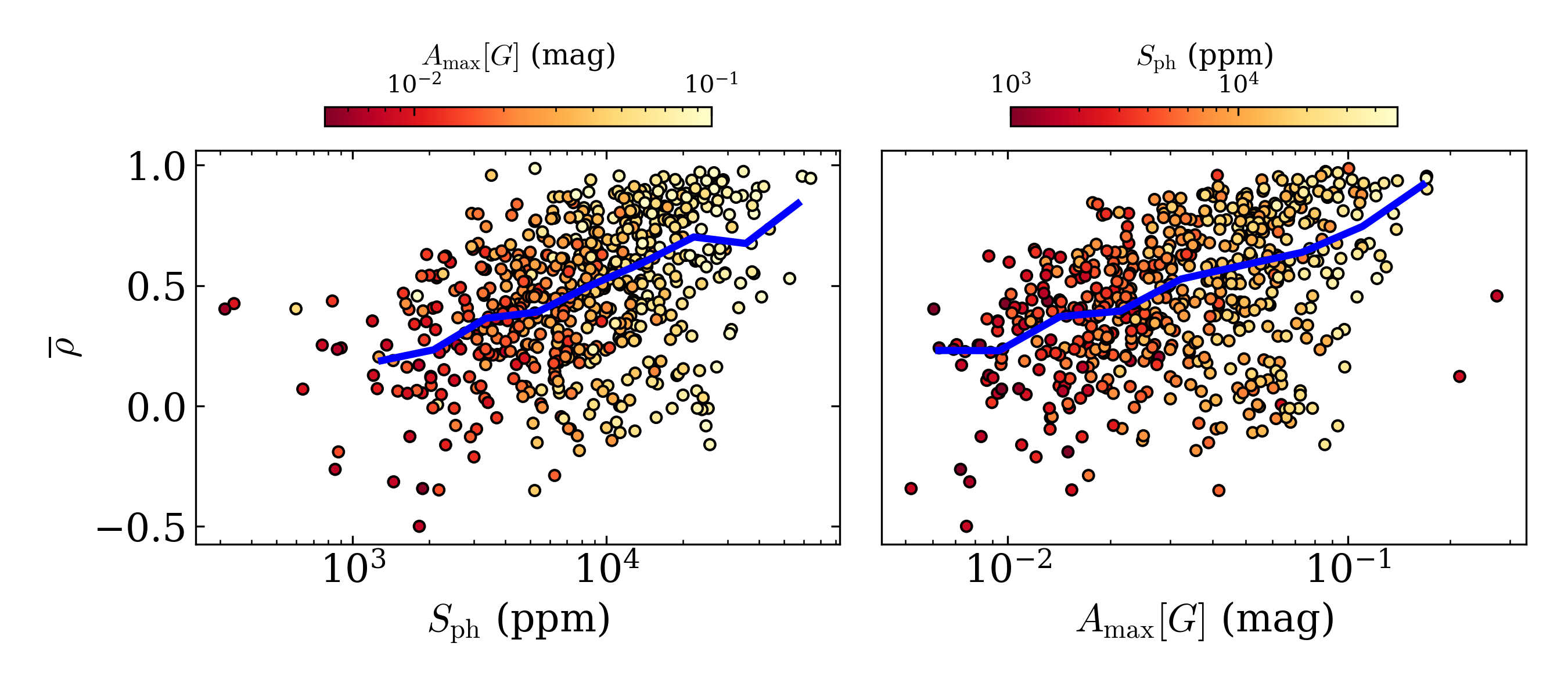}
    \caption{\textit{Left:} $\overline{\rho}$ vs $S_{\rm ph}$ diagram for the stars where the rotation period measured between K2 and \textit{Gaia} is cross-validated. $A_{\rm max}[G]$ is colour-coded. 
    {\color{black} The median value of the binned distribution along $A_{\rm max}[G]$ is shown in blue.}
    \textit{Right:} $\overline{\rho}$ vs $A_{\rm max}[G]$ diagram for the stars where the rotation period measured between K2 and \textit{Gaia} is cross-validated. $S_{\rm ph}$ is colour-coded.
    {\color{black} The median value of the binned distribution along $S_{\rm ph}$ is shown in blue.}    }
    \label{fig:activity_vs_rho}
\end{figure*}

Both K2 and \textit{Gaia} rotation measurements allows deriving photometric activity proxies. The K2 activity proxy that we use in this work is the $S_\mathrm{ph}$ defined in Sect.~\ref{sec:rotation_analysis} and already shown in the $G_{\rm BP} - G_{\rm RP}$ vs $P_\mathrm{rot}$ from Fig.~\ref{fig:bp_rp_prot_diagram_combined.png}.
For a \textit{Gaia} star with $n$ observed segments, the \textit{Gaia} activity proxy in the $G$ band is defined as 
\begin{equation}
    A_{\rm max} [G] = \max\limits_{i \in n} (G_{95\mathrm{th}, i} - G_{5\mathrm{th}, i}) \; ,
\end{equation}
where $G_{95\mathrm{th}, i}$ and $G_{5\mathrm{th}, i}$ are the 95th and the 5th percentile of the G magnitude distribution in the $i$th segment.

\begin{figure}[ht!]
    \centering
    \includegraphics[width=0.48\textwidth]{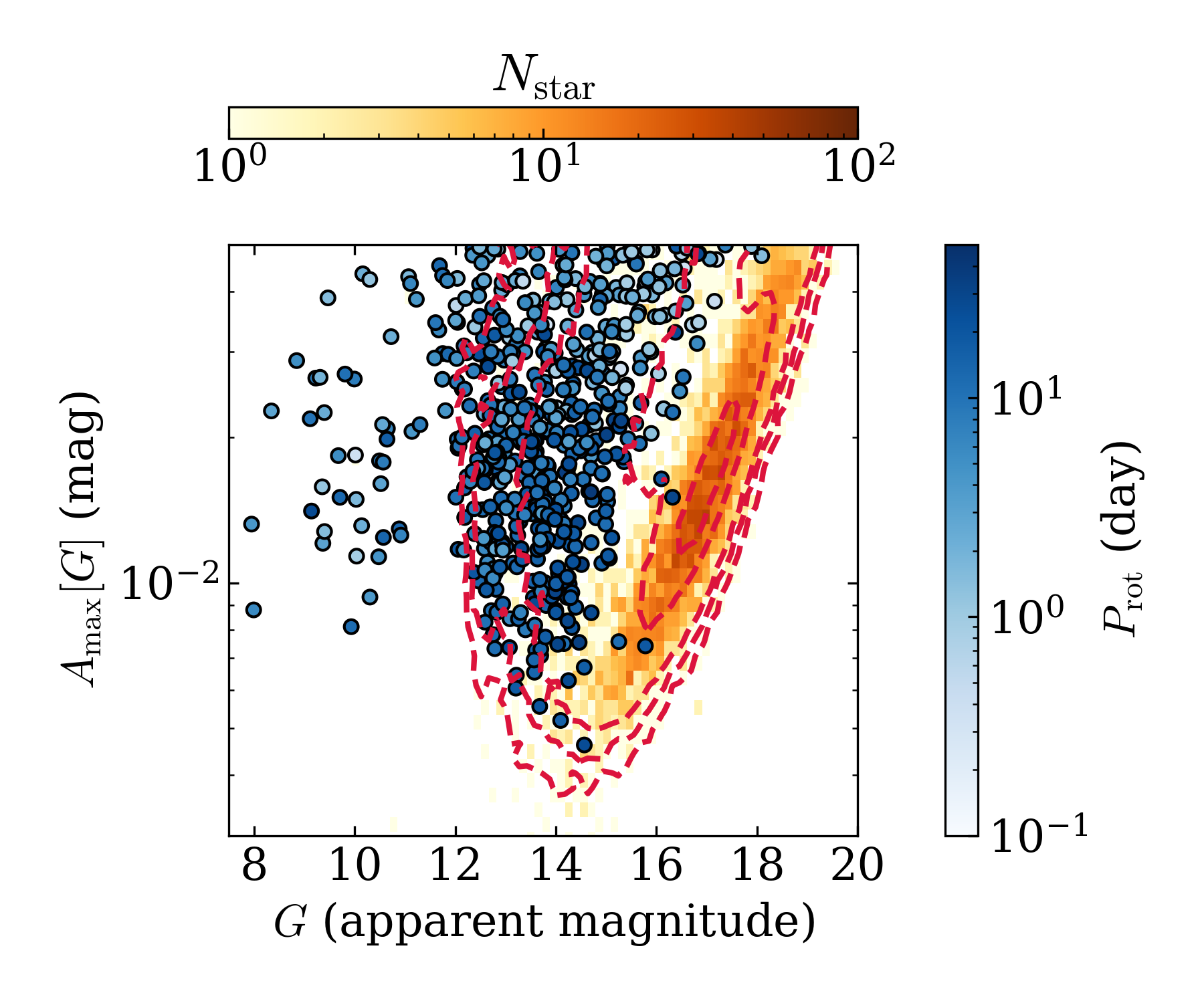}
    \caption{$A_{\rm max}[G]$ vs $G$ diagram for K2 stars (\textit{dots}) with $P_{\rm rot}$ colour-coded, compared with the density map of \textit{Gaia} DR3 low-activity UFR stars located at absolute ecliptic latitude lower than 10$\rm ^o$. The density contours of the full \textit{Gaia} DR3 rotation catalogue is represented for comparison in dashed red.}
    \label{fig:k2_and_ufr_amax}
\end{figure}

We can see in Fig.~\ref{fig:kepler_distribution} that most of the stars in our sample are located on the active edge of the \textit{Kepler} distribution, while on the contrary, it is visible from Fig.~\ref{fig:gaia_distribution} that they sample uniformly the \textit{Gaia} distribution, with the notable exception of the region where the \textit{Gaia} low-activity UFR are concentrated \citep[see][]{Lanzafame2019}, and the upper right edge (slow and active rotators), which are also depleted. 
The absence of low-activity UFR in K2 can be explained considering the relationship between the apparent magnitude $G$ and $A_{\rm max}[G]$. Indeed, as $G$ increases, rotation modulation detection of small amplitudes becomes more difficult with K2, as illustrated in Fig.~\ref{fig:k2_and_ufr_amax}. When we compare the $A_{\rm max}[G]$ vs $G$ distribution of the K2 sample with the \textit{Gaia} DR3 low-activity UFR sample, selected using the criterion defined by \citet{Lanzafame2019}, that is $P_{\rm rot} < 0.5$ and $A_{\rm max}[G] < 0.05$. Considering stars located at and absolute ecliptic latitude lower than 10$\rm ^o$ (that is roughly the range of latitude explored by the K2 mission), we see that there is almost no overlap between the two populations.
{\color{black}Only 9 stars in our cross-validated sample are inside the area defined by the \citet{Lanzafame2019} criterion, and most of them are on the high-activity edge of the UFR distribution (see Fig.~\ref{fig:gaia_distribution}).
For the rest,} we note that {\color{black}most of the} K2 stars that are located close to the bulk area of the UFR in the $A_{\rm max}[G]$ vs $G$ diagram are not fast rotators.
{\color{black}
This suggests that the UFR correspond to a population of stars that is globally located further away from us than the objects observed by \textit{Kepler} and K2.}

Also it should be noted that, given the selection cutoff operated for \textit{Gaia}, the post-Kraft break F-type stars with fast rotation and low activity, visible in the lower left part of Fig.~\ref{fig:kepler_distribution} \citep[see][]{Santos2021}, are not represented in our sample, as already noted by \citet{Distefano2023}. The correlation between $S_{\rm ph}$ and $A_{\rm max} [G]$ is already apparent in these two diagrams. 

In Fig.~\ref{fig:sph_vs_max_g}, we directly compare the $S_\mathrm{ph}$ index to $A_{\rm max}[G]$.
Considering the {\color{black}Pearson} correlation coefficients $\rho$\footnote{\color{black}We underline the fact that the activity index $A_{\rm max}[G]$ and the correlation coefficients $\rho$ are provided in the \textit{Gaia} database only for objects of the \texttt{vari\_rotation\_modulation} table.} between $G$ magnitude variation and $G_{\rm BP} - G_{\rm RP}$ colour provided segment-wise in the \textit{Gaia} rotation catalogue, we also compute a mean correlation coefficient $\overline{\rho}$ that we show together with the two activity indexes. 
Unavailable from mono-band photometric observations such as those performed by \textit{Kepler}/K2, this coefficient therefore quantifies the level of correlation between the variation of the stellar apparent magnitude and the stellar colour. 
In other words, a positive coefficient means that an apparent darkening of the star (e.g. when star spots appear in the field of view) goes with a reddening of its emitted spectrum, while a star with a negative coefficient becomes more blue when darkening. The correlation case can be schematically interpreted as the passage of a cold dark spot which results in a simultaneous darkening and reddening of the stellar disc. 
A strong anti-correlation suggests on the contrary that the modulation is connected to the orbit of a hotter eclipsing companion.
The $\overline{\rho}$ coefficient is positive for the vast majority of the stars, only a few of them exhibit an anti-correlated behaviour. 
An important property that we highlight is that $\overline{\rho}$ tends to increase with $S_\mathrm{ph}$ index and $A_{\rm max}[G]$, as visible in Fig.~\ref{fig:activity_vs_rho}, where we directly compare $S_{\rm ph}$ and $A_{\rm max}[G]$ with $\overline{\rho}$. Interestingly, for a small fraction of active stars, we have $\overline{\rho} < 0$.

\subsection{Targets with Gaia spectroscopic characterisation}

\begin{figure}[ht!]
    \centering
    \includegraphics[width=0.49\textwidth]{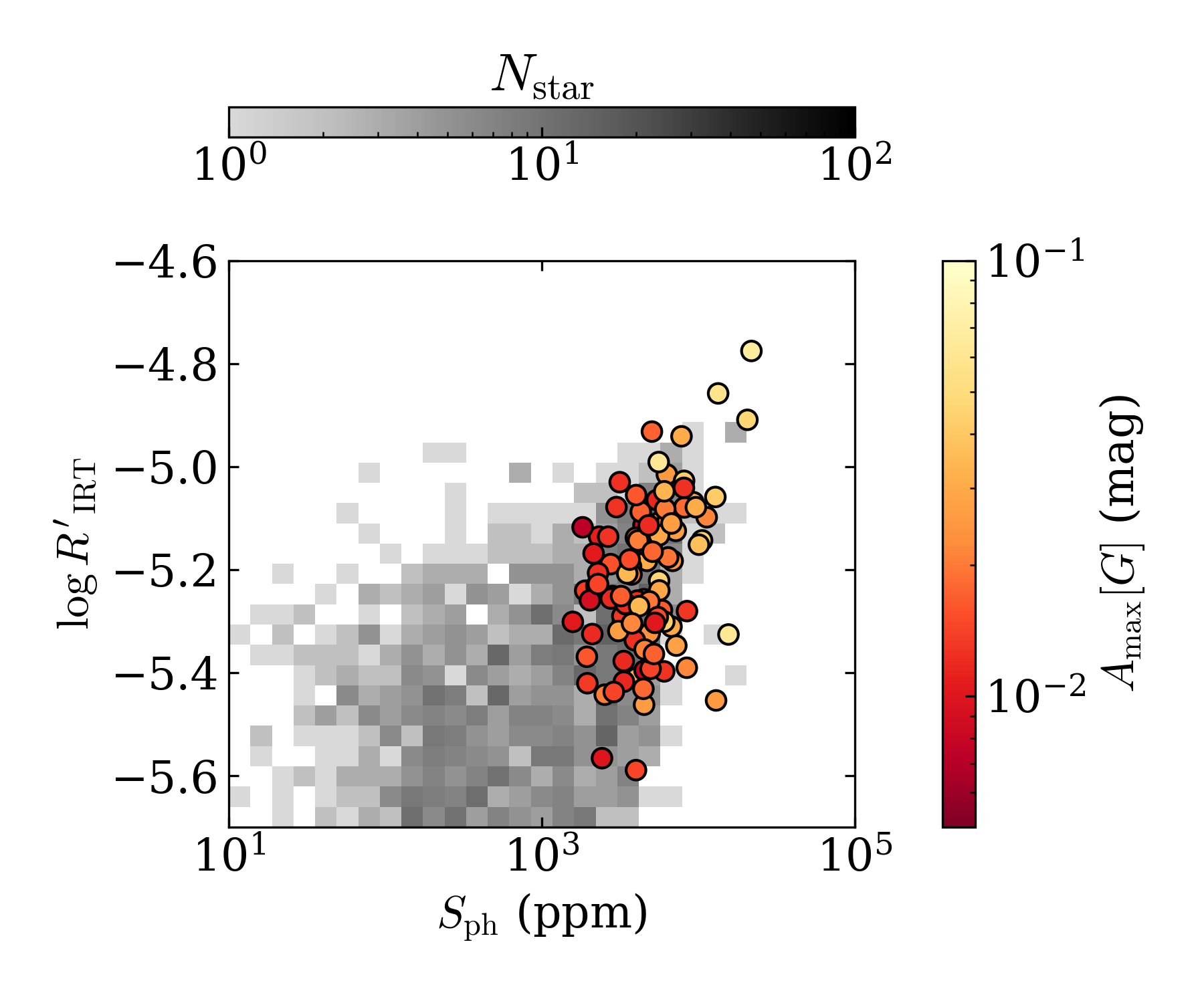}
    \caption{$\log R'_{\rm IRT}$ vs $S_\mathrm{ph}$ diagram with $A_{\rm max} [G]$ colour-coded for all stars with a reliable K2 rotation measurement where $\log R'_{\rm IRT}$ could be computed. The distribution of \textit{Kepler} stars from \citet{Santos2019,Santos2021} for which $\log R'_{\rm IRT}$ could be computed is shown for comparison.}
    \label{fig:sph_vs_log_irt_full_good_rot}
\end{figure}

\begin{figure}[ht!]
    \centering
    \includegraphics[width=0.49\textwidth]{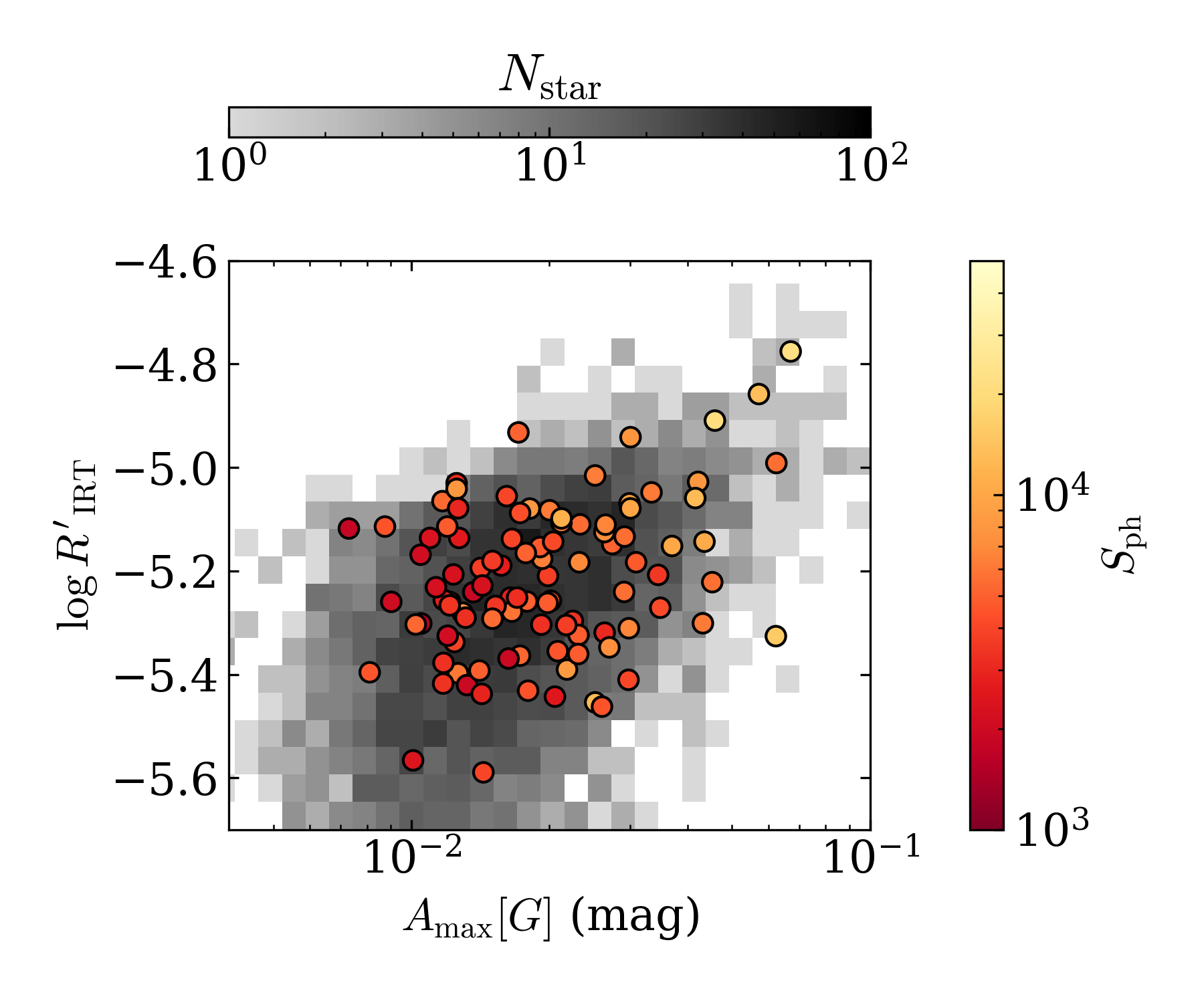}
    \caption{$\log R'_{\rm IRT}$ vs $A_{\rm max} [G]$ diagram with $S_\mathrm{ph}$ colour-coded for all stars with a reliable K2 rotation measurement where $\log R'_{\rm IRT}$ could be computed. The distribution of \textit{Gaia} DR3 stars with rotation measurements for which $\log R'_{\rm IRT}$ could be computed is shown for comparison.}
    \label{fig:amax_gaia_vs_log_irt_full_good_rot}
\end{figure}

In this section, we now consider all the stars of our sample for which we were able to measure a reliable rotation period from the K2 EVEREST light curve and we turn to the possibilities of characterisation offered by \textit{Gaia} infrared spectroscopic observations.
Using the RVS data, \citet{Lanzafame2023} derived a chromospheric activity index $\alpha_{\rm Ca \, II - IRT}$. The $\alpha_{\rm Ca \, II - IRT}$ being relevant only to compare stars with similar global properties, \citet{Lanzafame2023} defined the $R'_{\rm IRT}$ activity indicator and provided the following formula to estimate it from the effective temperature, $T_\mathrm{eff}$, the metallicity, $\rm [M/H]$, and $\alpha_{\rm Ca \, II - IRT}$.
\begin{equation}
    \log R'_{\rm IRT} = c_0 + c_1 \theta + c_2 \theta^2 + c_3 \theta^3 + \log \alpha_{\rm Ca \, II - IRT} \; ,
\end{equation}
where $\theta = \log T_\mathrm{eff}$, and the logarithms are decimal logarithms. The $c_i$ coefficients depend $\rm [M/H]$ and are interpolated using Table~1 from \citet{Lanzafame2023}. 
Given the bounds of this table, we compute $\log R'_{\rm IRT}$ only for stars within the range $-0.5 < [\rm M/H] < 0.5$.
We also remove stars for which $\alpha_{\rm Ca \, II - IRT}$ is smaller than three times its uncertainties, as the $\alpha_{\rm Ca \, II - IRT}$ measurement in this case is probably dominated by random noise, and those who have $\log R'_{\rm IRT} < -5.7$, as these stars can be considered inactive.
To obtain an homogeneous sample, we consider only stars for which a \textit{Gaia} RVS spectroscopic $T_\mathrm{eff}$ and $\rm [M/H]$ values are available.

We directly compare $\log R'_{\rm IRT}$ to $S_{\rm ph}$ and $A_{\rm max}[G]$ in Fig.~\ref{fig:sph_vs_log_irt_full_good_rot} and \ref{fig:amax_gaia_vs_log_irt_full_good_rot}, respectively.
Although $S_{\rm ph}$ and $A_{\rm max}[G]$ are both correlated with $\log R'_{\rm IRT}$, there is a strong dispersion in the relation between the parameters. We remind that photometric activity indicators such as $S_{\rm ph}$ and $A_{\rm max}[G]$ are only lower-limit proxy of activity as they might be biased by the inclination of the stellar rotation axis with respect to the observer line-of-sight, as well as the latitude of emergence of the active regions. We also verify again that the stars in our sample coincide with the active edge (larger $\log R'_{\rm IRT}$) of the \textit{Kepler} distribution.

\section{Discussion \label{sec:discussion}}

\subsection{Discrepancies between \textit{Gaia} and K2 \label{sec:inconsistence_k2_gaia}}

\begin{figure}[ht!]
    \centering
    \includegraphics[width=0.49\textwidth]{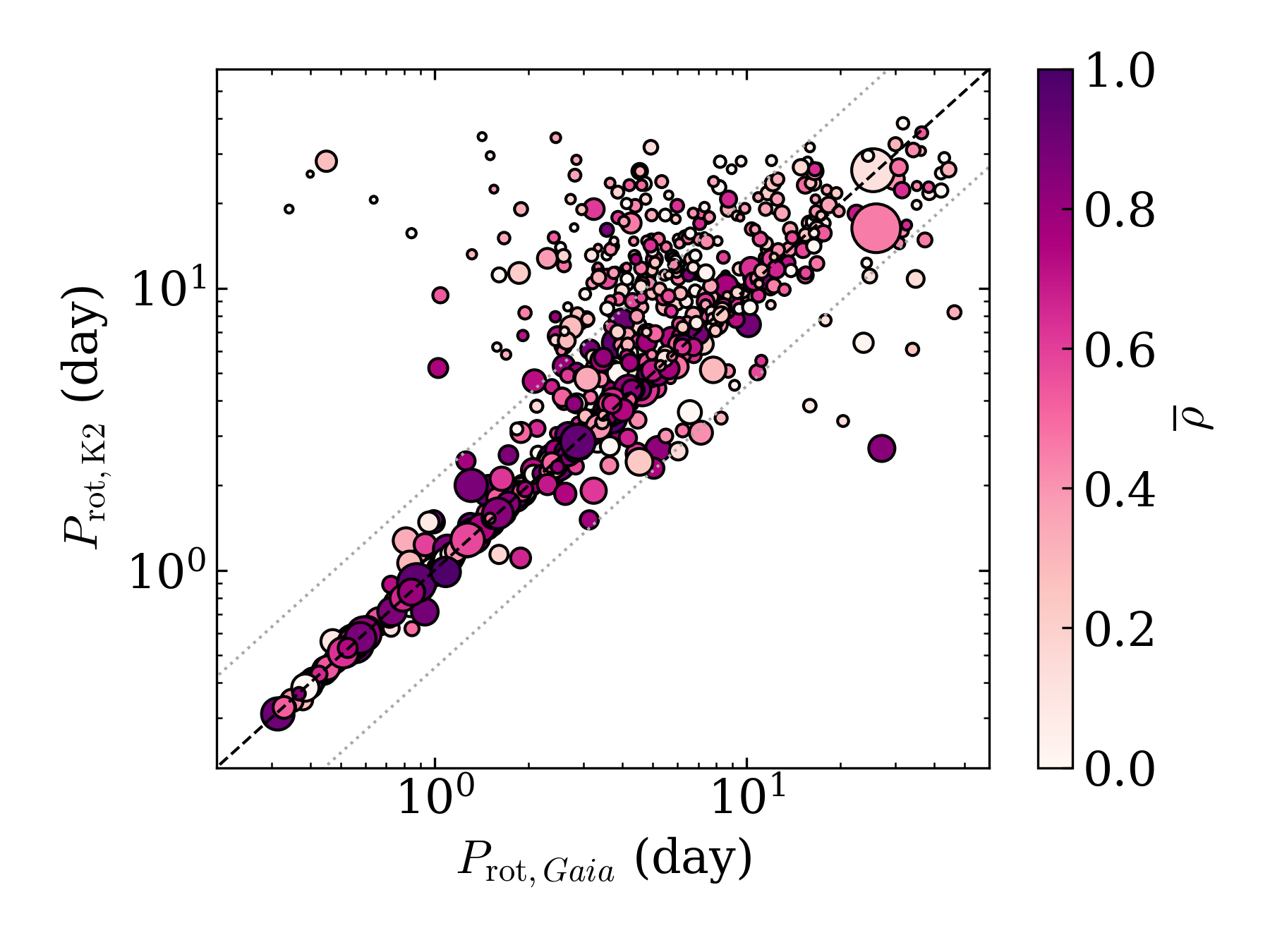}
    \caption{Comparison between the rotation periods measured and validated in the K2 light curves, $P_\mathrm{rot,K2}$, and the reference values from the \textit{Gaia} DR2 and DR3 catalogues, $P_{\mathrm{rot},Gaia}$. 
    The 1:0.45, 1:2.1 (dotted grey lines), and 1:1 (dashed black line) lines are shown. 
    The mean correlation coefficient $\overline{\rho}$ is color-coded, and the dot size is proportional to the $A_{\rm max} [G]$ index.
    For readability, $\overline{\rho}$ values are color-coded only on the 0 to 1 range. 
    }
    \label{fig:prot_gaia_k2}
\end{figure}

\begin{figure*}[ht!]
    \centering
    \includegraphics[width=\textwidth]{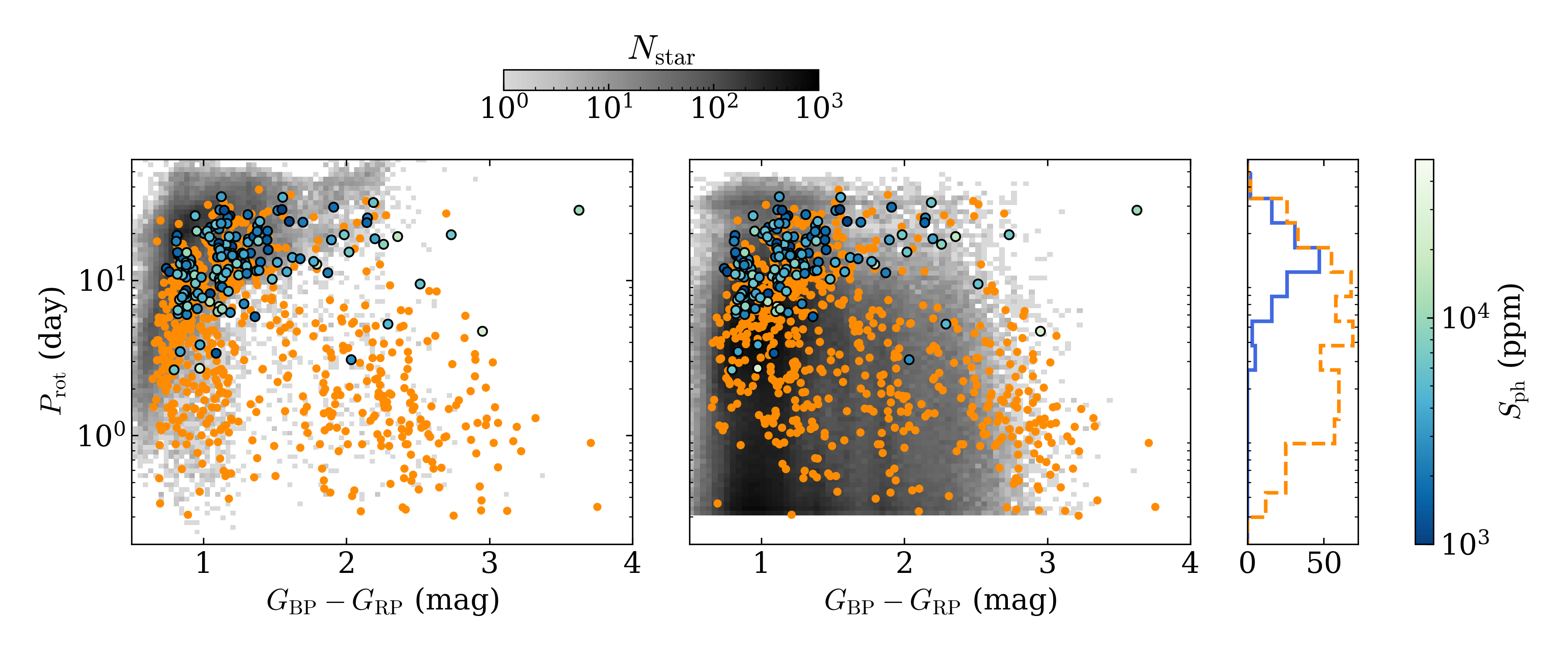}
    \caption{$P_\mathrm{rot}$ vs $G_{\rm BP} - G_{\rm RP}$ diagram. The $S_{\rm ph}$ value is colour-coded for the stars for which \textit{Gaia} and K2 rotation measurements are cross-validated. 
    The stars for which the measurements are consistent are shown in orange.
    On the left panel, the sample is compared with the density distribution of the \textit{Kepler} stars from \citet{Santos2019,Santos2021} (in grey) while on the right panel it is compared with the density distribution of the stars from \textit{Gaia} DR3 (in grey).}
    \label{fig:bp_rp_prot_diagram_combined_inconsistent}
\end{figure*}

As stated in Sect.~\ref{sec:rotation_period_k2_gaia}, while we found consistent measurements for more than 80\% of the stars where we validate the rotation measurements in the K2 light curve, the rotation periods obtained from \textit{Gaia} observations and K2 light curves differ for a fraction of this sample.

Before exploring in more depths the origins of these differences, it is interesting to summarise and briefly discuss the respective advantages of \textit{Gaia} and \textit{Kepler}/K2 when it comes to measuring stellar surface rotation.
On the one hand, the \textit{Kepler}/K2 standard cadence of 30 minutes is well suited for surface activity modulation follow-up, but while most \textit{Kepler} targets have a 4-year long light curve, the design of the K2 mission has for consequence that K2 light curves are 90-day long segments, which in turn limits the possibility to monitor rotation in slow rotators.
On the other hand, \textit{Gaia} observations are sparse but cover several years, the high photometric precision of the observation making them ideal to monitor long-term activity-induced modulation of the observed stars. 
In addition, \textit{Gaia} pixel size is several order of magnitudes smaller (59 milli-arcseconds in the scanning direction) than for \textit{Kepler} (3.98 arcseconds per pixel), which is important in the case of crowded fields, where light curves can be contaminated by neighbouring stars of similar brightness. \textit{Gaia} sensitivity also allows it to observe fainter stars (as already mentioned in Sect.~\ref{sec:rotation_period_k2_gaia} and Fig.~\ref{fig:bp_rp_relative_gmag_diagram}), it has also access to multi-colour photometry while \textit{Kepler}/K2 is mono-band. 
Finally, we can mention that the high-cadence of \textit{Kepler}/K2 is necessary in order to perform more in-depth studies of the morphology of the active regions \citep[e.g. starspot modelling, see][]{Lanza2016,Breton2024spotmodelling}.

We remind that rotation periods were measured in \textit{Gaia} using the following methodology \citep{Lanzafame2018,Distefano2023}: for a given star, in each time-series segment obtained by the instrument, the highest peak of the Lomb-Scargle periodogram \citep{Lomb1976,Scargle1982,Zechmeister2009} is selected and validated if its false alarm probability is below 0.05. In this case, a sinusoidal model is fitted on the data in order to estimate the amplitude of the modulation. Finally, considering the set of periods extracted from the different segment, the mode of the distribution is taken as the best rotation period \citep[see][for more details]{Lanzafame2018}. 
We remind that each segments contains at least 12 \textit{Gaia} measurements and lasts no longer than 120 days \citep{Lanzafame2018}.

In order to investigate the possible origins of these discrepancies, we compare in Fig.~\ref{fig:prot_gaia_k2} the rotation periods we extract from the K2 light curves to the reference \textit{Gaia} periods from DR2 and DR3. 
On the one hand, the diagram shows that, up to $P_{\rm rot} \sim 3-4$~days, the overall agreement between K2 and \textit{Gaia} is very good, validating that \textit{Gaia} is extremely efficient at recovering rotation periods for fast rotators. 
On the other hand, we note that, for the inconsistent measurements, \textit{Gaia} tends to underestimate the stellar rotation period in the majority of the cases, although in some cases the \textit{Gaia} rotation period is overestimated. 

It is important to note that stars outside the cross-validated sample exhibit in general a small activity index $A_{\rm max}[G]$ and a small correlation coefficient $\overline{\rho}$. 
We also consider the stability criterion defined by \citet{Distefano2023}: a star is flagged as having a stable modulation if the \textit{Gaia} best rotation period differs from less than 5\% from the main periodicity in the full \textit{Gaia} timeseries. We note that 57\% of the stars in the cross-validated sample fulfill this stability criterion, while it is the case of only 47\% of the other sample.
This suggest that these different parameters may be used as quality assessments quantities. 
Another aspect to keep in mind is the fact that the K2 stars we deal with have been observed by \textit{Gaia} in a maximum of six segments.
At different ecliptic latitudes, especially around 45$\rm ^o$ and close to the ecliptic pole, stars have been observed during a significantly larger number of visits \citep[see][]{Distefano2023}.
We also compare in Fig.~\ref{fig:bp_rp_prot_diagram_combined_inconsistent} the location of the two populations on the $P_\mathrm{rot}$ vs $G_{\rm BP} - G_{\rm RP}$ diagram, with the inferred $S_{\rm ph}$ as colour-coding. Again, in this diagram, we see that the distribution of the two subsamples is quite different, with the stars with inconsistent measurements concentrated on the bulk of the \textit{Kepler} distribution.  

\subsection{\color{black}Identifying \textit{Gaia} stars similar to the K2 subsample}

\begin{figure}[ht!]
    \centering
    \includegraphics[width=\linewidth]{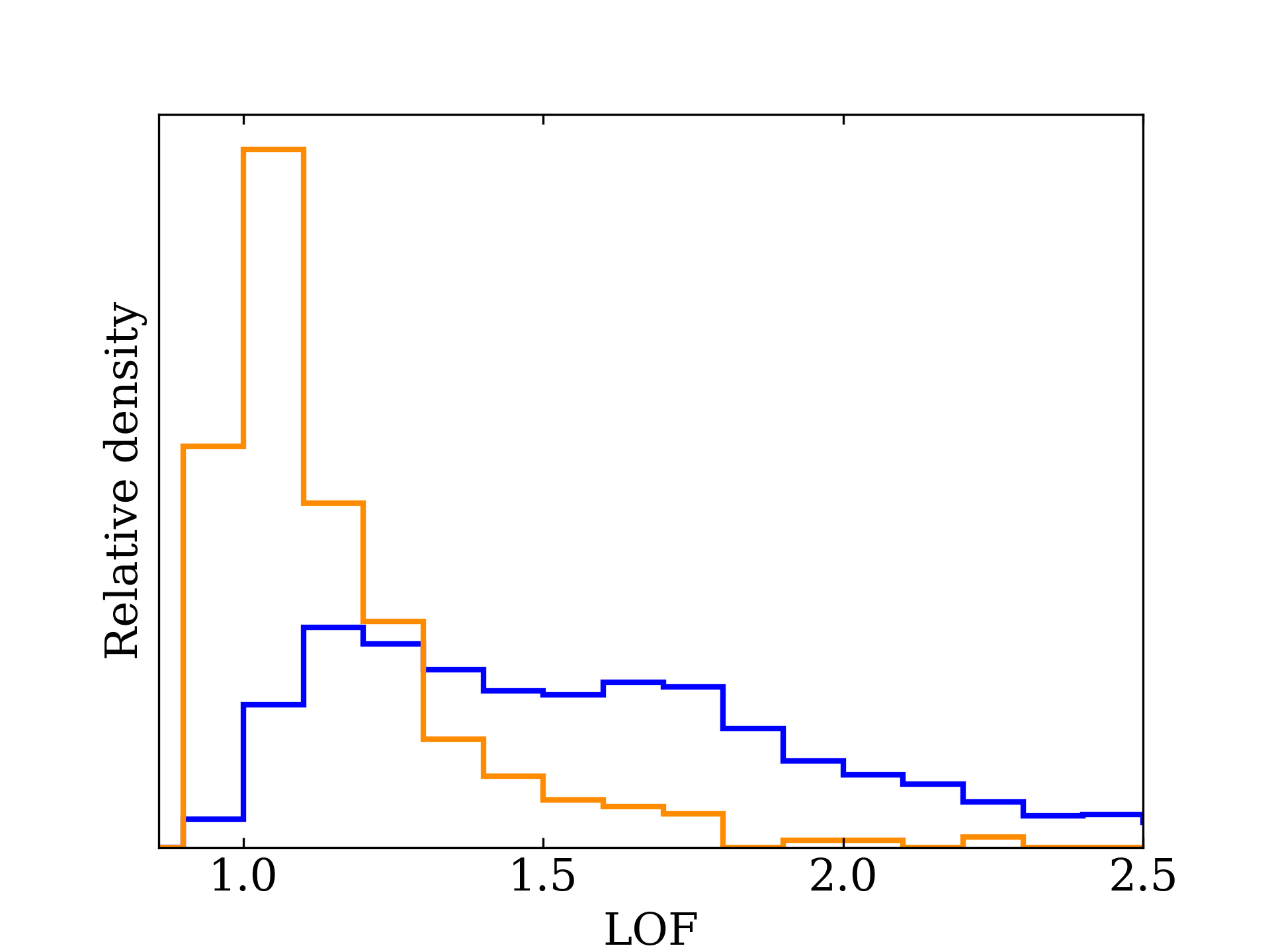}
    \caption{\color{black}LOF computed for our K2 subsample (orange) and for the full set of \textit{Gaia} DR3 stars (blue).}
    \label{fig:lof_histogram}
\end{figure}

\begin{figure}[ht!]
    \centering
    \includegraphics[width=\linewidth]{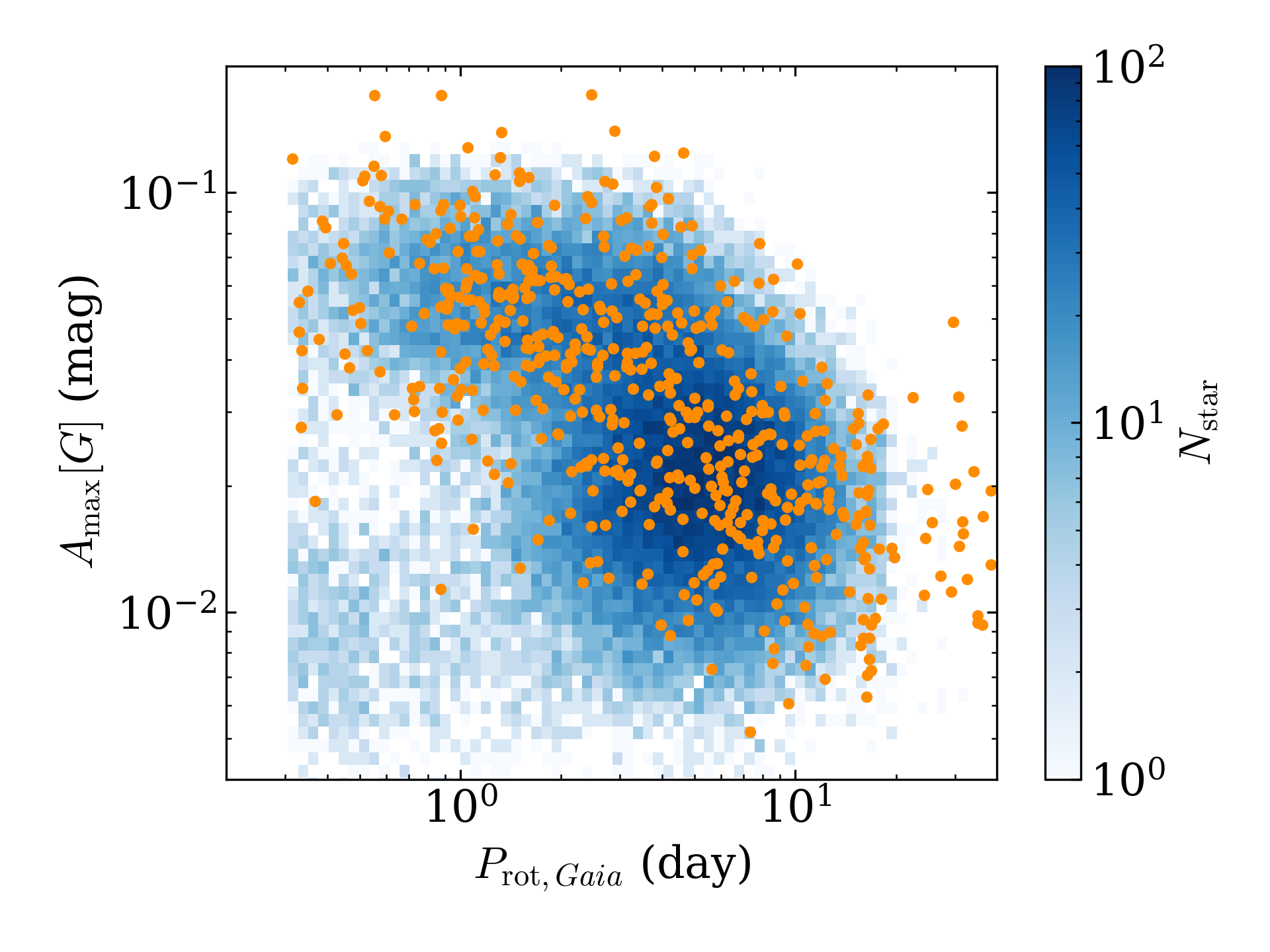}
    \caption{$A_{\rm max}[G]$ vs $P_{\mathrm{rot}, Gaia}$ diagram for \textit{Gaia} DR3 stars {\color{black} which have a LOF $\leq 1.1$}. The density of stars is color-coded. {\color{black}The K2 cross-validated sample is shown in orange for comparison.}}
    \label{fig:lof_selected}
\end{figure}

{
\color{black}

In this section, we investigate the possibility to select \textit{Gaia} stars that have properties similar to the ones in the K2 cross-validated sample. To do so, we use the Local Outlier Factor \citep[LOF,][]{Breunig2000}, with the implementation provided by \texttt{scikit-learn} \citep{scikit-learn}.
The LOF is an unsupervised method that is able to measure the local density of element clusters in a dataset, and, from this, to estimate the degree of abnormality of unknown elements. 
A data point with a LOF close to 1 can be interpreted as a probable inlier while a higher value suggests that the element is an outlier. It should be noted that there is no strict threshold distinguishing inliers from outliers and that the cutoff choice to separate inliers from outliers is therefore arbitrary. It is also important to have in mind that the method allows obtaining the LOF of each element of the training dataset, and that these LOF are not mandatorily close to 1, depending on the position of a given elements with respect to the density distribution of the dataset.

The parameters we use to compute the LOF are the following: \textit{Gaia} rotation period and corresponding error, $A_{\rm max}[G]$ activity index, number of observed segments, mean correlation coefficient, periodicity stability criterion, larger and lower value of the Lomb-Scargle false alarm probability estimated for the observation segments, magnitude $G$, and $G_{\rm BP} - G_{\rm RP}$.
In Fig.~\ref{fig:lof_histogram}, we compare the LOF distribution obtained for the reference K2 cross-validated sample and the LOF distribution of the other stars in the \textit{Gaia} DR3 rotation catalogue. We note that, while most of the stars in the reference sample have a LOF close to 1, there exist a tail of elements with a higher LOF, and therefore a higher degree of abnormality in the cross-validated sample. Additionally, only a small fraction of the \textit{Gaia} DR3 rotation catalogues targets have a LOF close to 1. Selecting only stars with LOF $\leq 1.1$, that is 40,423 objects, we represent them in the $A_{\rm max}[G]$ vs $P_{\mathrm{rot},Gaia}$ of Fig.~\ref{fig:lof_selected}. Although some of the selected stars are fast rotators with low level of activity (some of them laying in the UFR region), and very few stars with $P_{\rm rot} > 20$~days have been selected, we note that the parameter space covered by the stars with LOF~ $\leq 1.1$ is globally located close to the bulk of the cross-validated K2 sample. As these values should be useful for future statistical studies, the full set of computed LOF is available, as summarised in Table~\ref{table:score_table}.

}

\begin{table*}[t]
\caption{Excerpt of the table summarising the properties of the stars in the \textit{Gaia} DR3 catalogue and their LOF.}
\label{table:score_table}
\centering
\begin{tabular}{ccccccc}
\hline\hline
\textit{Gaia} DR3 source ID & $G$ (mag) & $G_{\rm BP} - G_{\rm RP}$ (mag) & $M_G$ (mag) & $P_\mathrm{rot}$ (day) & $A_\mathrm{max}[G]$ (mag) & LOF \\
\hline
11964580592929024 & 14.7 & 1.24 & 6.69 & 0.99 $\pm$ 0.00 & 0.011 & 1.80 \\
11980420432272896 & 13.3 & 1.36 & 6.14 & 7.25 $\pm$ 0.00 & 0.040 & 0.98 \\
11981760462138496 & 17.4 & 1.31 & 7.78 & 0.43 $\pm$ 0.00 & 0.022 & 1.64 \\
11987807776011392 & 14.6 & 1.07 & 5.00 & 6.06 $\pm$ 0.29 & 0.010 & 1.00 \\
11987807776011520 & 15.1 & 1.07 & 5.43 & 2.04 $\pm$ 0.03 & 0.051 & 1.02 \\
... & ... & ... & ... & ... & ... & ... \\
\hline
\end{tabular}
\tablefoot{
A machine-readable version of this table is available. When possible, the $G_{\rm BP}-G_{\rm RP}$ magnitudes are corrected with the Apsis dereddening coefficients.
}
\end{table*}

\section{Conclusion \label{sec:conclusion}}

In this work, we combined \textit{Gaia} and K2 observations to characterise the sample of rotating solar-type stars that were observed by both missions. The rotation periods we extracted from the EVEREST reduction of the K2 observations were globally in good agreement with the \textit{Gaia} measurements, especially for the fast rotators ($P_{\rm rot}$ below 3-4~days). We discussed the discrepancies we observed between \textit{Gaia} and K2 for a fraction of our sample, noting that these stars were slow rotators with small level of activity, for which photometric rotation measurements are notoriously difficult to carry out.
We also reminded that the limited number of segments available for \textit{Gaia} at the ecliptic latitudes where K2 stars are located. 
We also noted that the population of \textit{Gaia} low-activity UFR discussed by \citet{Lanzafame2019} is not represented in our K2 sample, preventing us to investigate the properties of this region of the diagram. 
In the context of this work, the elusive character of the UFR is explained by the fact that their $A_{\rm max}[G]$ vs $G$ relationship makes the vast majority of them inaccessible for the instrumental capabilities of K2 (see Fig.~\ref{fig:k2_and_ufr_amax}).
We proceeded to an extensive comparison of the photometric activity indicators obtained by each missions, we then considered the cases of the targets for which \textit{Gaia} RVS observations are available, which allowed us to run a comparison between the $\log R'_{\rm IRT}$ indicator defined by \citep{Lanzafame2023} and the $S_{\rm ph}$ photometric indicator, for both \textit{Kepler} and K2 stars.
Finally, noting that, in our sample, the stars with consistent K2 and \textit{Gaia} parameters exhibit distinct properties in term of variability amplitude and correlation between the different \textit{Gaia} colors, {\color{black} we computed the LOF of every star in the \textit{Gaia} DR3 rotation catalogues with the K2 cross-validated sample as reference. We showed that the rotation/activity distribution of the 40,423 stars with LOF~$\leq 1.1$ exhibited a good similarity to the cross-validated K2 sample, which make them a good subsample to consider for future statistical studies.}

Concerning other space missions, we anticipate that, due to their common strategy of full scale coverage, TESS and \textit{Gaia} should have an important number of stars with measured rotation in common. However, we remind that the 27-day sector length of TESS observation means that it is difficult to measure the rotation periods of moderately slow and slow rotators located outside of the TESS continuous viewing zone.  
In this sense, the possibility to work with long temporal baseline continuous observations such as those that will be acquired by PLATO is crucial. 
The intersection between the two long-pointing PLATO fields of observation and \textit{Gaia} targets with measured rotation periods should remain limited \citep[see Fig.~2 from][compared to our Fig.~\ref{fig:galactic_location}]{Nascimbeni2022} but a fraction of the PLATO stars should still possess previous \textit{Gaia} rotation measurements. Nevertheless, future data releases from \textit{Gaia} should provide an increased set of activity indicators and rotation measurements for stars in the PLATO fields. 

\begin{acknowledgements}
The authors want to thank the anonymous referee for helpful comments that allowed improving the manuscript.
S.N.B acknowledges support from PLATO ASI-INAF agreement no. 2022-28-HH.0 "PLATO Fase D".
D.B.P acknowledges support from PLATO CNES grant.
S.N.B. thanks A.F.~Lanza and R.A.~García for useful discussions and advices. 
This paper includes data collected by the \textit{Kepler} mission and obtained from the MAST data archive at the Space Telescope Science Institute (STScI). Funding for the Kepler mission is provided by the NASA Science Mission Directorate. STScI is operated by the Association of Universities for Research in Astronomy, Inc., under NASA contract NAS 5–26555.
This work has made use of data from the European Space Agency (ESA)
mission {\it Gaia} (\url{https://www.cosmos.esa.int/gaia}), processed by
the {\it Gaia} Data Processing and Analysis Consortium (DPAC,
\url{https://www.cosmos.esa.int/web/gaia/dpac/consortium}). Funding
for the DPAC has been provided by national institutions, in particular
the institutions participating in the {\it Gaia} Multilateral Agreement.
This work made use of the \texttt{gaia-kepler.fun} crossmatch database created by Megan Bedell. This research made use of Lightkurve, a Python package for Kepler and TESS data analysis.
\\
\textit{Software:} 
\texttt{star-privateer} \citep{Breton2024PLATO},
\texttt{numpy} \citep{harris2020array}, \texttt{matplotlib} \citep{Hunter:2007}, \texttt{scipy} \citep{2020SciPy-NMeth}, 
\texttt{astropy} \citep{astropy:2022}, 
\texttt{astroquery} \citep{Ginsburg2019astroquery},
\texttt{pandas} \citep{Pandas2020},
\texttt{lightkurve} \citep{Lightkurve2018},
\texttt{scikit-learn} \citep{scikit-learn}

\\
\end{acknowledgements}

\bibliographystyle{aa} 
\bibliography{biblio.bib} 

\appendix

\section{Targets belonging to open clusters \label{sec:open_clusters}}

\begin{figure}[ht!]
    \centering
    \includegraphics[width=0.49\textwidth]{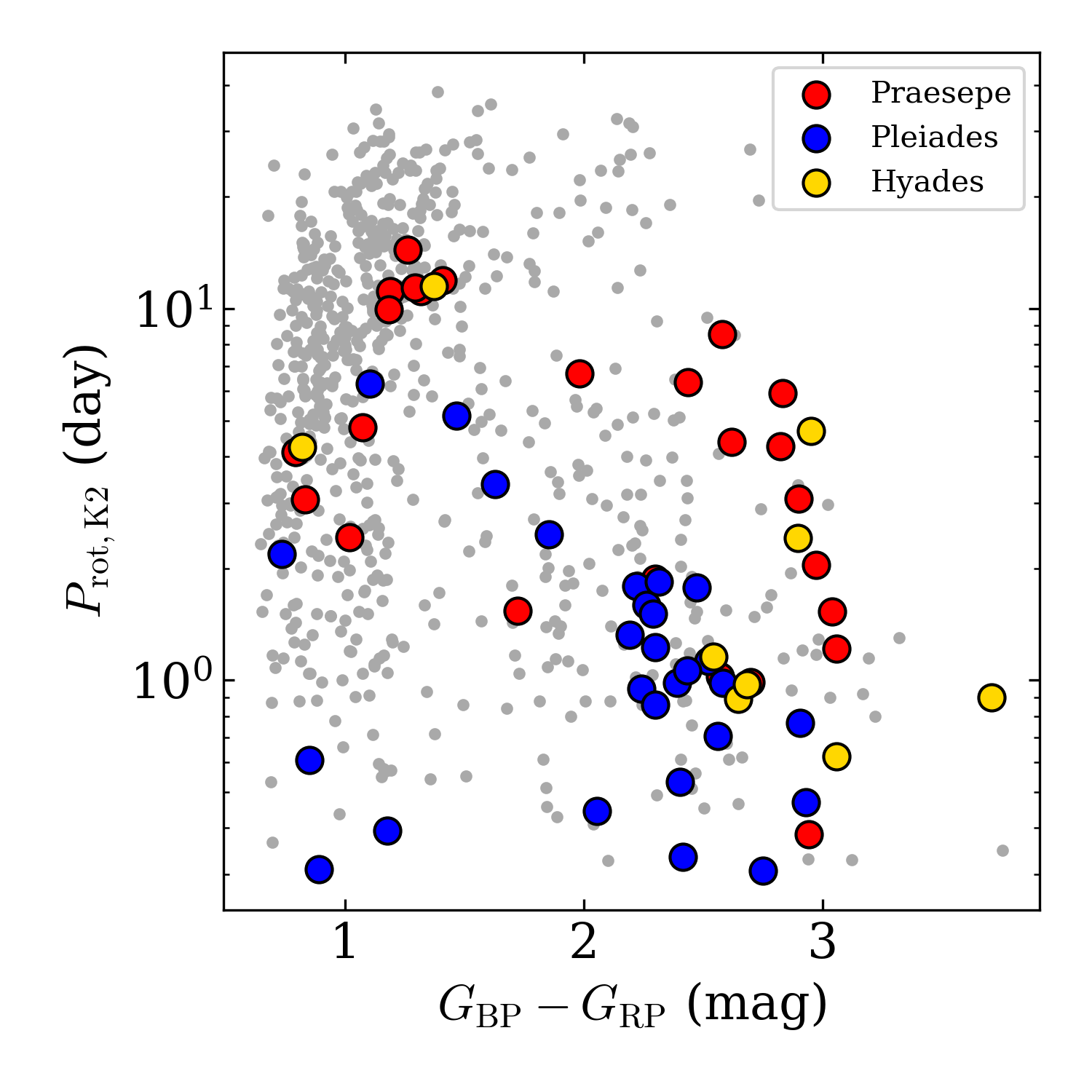}
    \caption{
     $P_{\rm rot}$ vs $G_{\rm BP} - G_{\rm RP}$ diagram for Praesepe (red), Pleiades (blue) and NGC~1647 (yellow) stars.
    The other stars of our sample are represented in grey for comparison.}
    \label{fig:clusters_bp_rp_prot}
\end{figure}

\begin{figure}[ht!]
    \centering
    \includegraphics[width=0.49\textwidth]{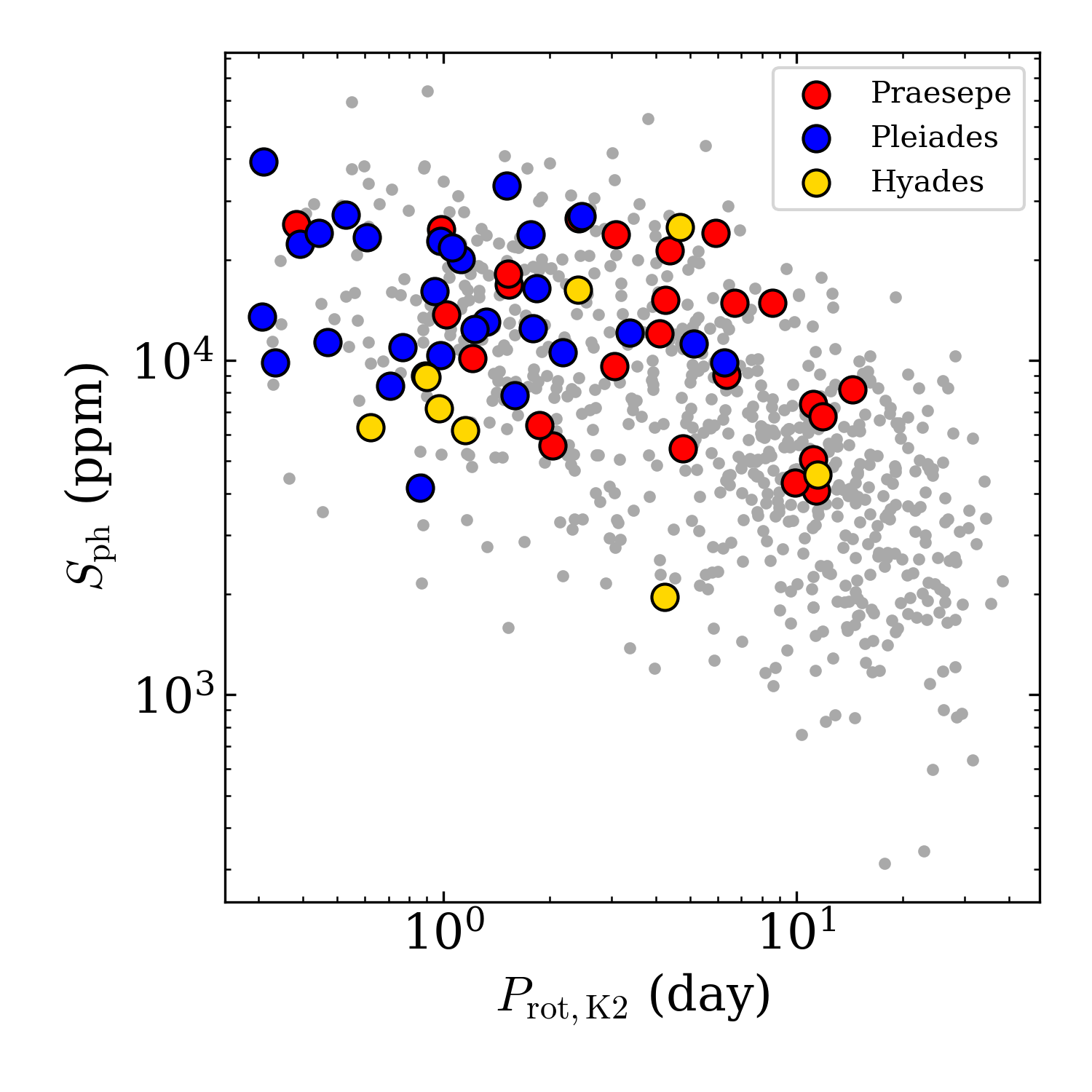}
    \caption{
    $S_{\rm ph}$ vs $P_{\rm rot}$ diagram for Praesepe (red), Pleiades (blue) and NGC~1647 (yellow) stars.
    The other stars of our sample are represented in grey for comparison.}
    \label{fig:clusters_prot_sph}
\end{figure}

By cross-matching our list of targets with the cluster membership catalogue from \citet{Hunt2023}, we find that 78 stars for which we were able to measure a rotation periods are part of known open clusters with a probability larger than 0.5. Among these stars, 28 belong to Pleiades, 25 to Praesepe, and 9 to Hyades. 
\citet{Rebull2016a,Rebull2016b,Stauffer2016} used K2 light curves to perform an extensive study of stellar variability in about one thousand candidate Pleiades members, reporting rotation measurements for 716 high-confidence members of the cluster.
\citet{Brown2021} showed that stars for Pleiades exhibited the so-called mid-frequency continuum (MFC), a signal of probable magnetic origin that significantly contributed to the variability of the targets where it was found, without being correlated with the modulation amplitude from active regions. 
Rotational properties in Praesepe were also already explored with K2 by \citet{Douglas2017}, who reported rotation measurements for 677 targets. 
More recently, rotation periods for stars in Pleiades, Praesepe, and Hyades were provided by \citet{Long2023}, among other open clusters. 
The age of Hyades is between 500 and 800 Myr \citep{PerezGarrido2018}, while for the Praesepe cluster it is thought to be between 580 and 700 Myr \citep{Gossage2018,Douglas2019}. 
We show in Fig.~\ref{fig:clusters_bp_rp_prot} to the $G_{\rm BP} - G_{\rm RP}$ versus $P_{\rm rot, K2}$ diagram. Praesepe and Hyades stars are globally slower rotators as expected from the age difference with Pleiades. 
The $P_{\rm rot, K2}$ versus $S_{\rm sph}$ diagram of Fig.~\ref{fig:clusters_prot_sph} shows that the slowest rotators of Praesepe and Hyades have transitioned towards less active regimes, while some of them still have an activity level comparable with the one observed in Pleiades.

\section{ROOSTER training set \label{sec:rooster_training_set}}

In this appendix, we describe how the ROOSTER training set was adapted in order to correctly predict the rotation periods of the K2 targets. The training set for this work is composed of \textit{Kepler} stars with rotation measurements performed by \citet{Santos2019,Santos2021}. 
We consider the KEPSEISMIC light curves \citep{Garcia2011,Garcia2014,Pires2015} where the signal beyond 55 day has been filtered out. 
Given the length of the K2 light curves and the prior knowledge on the period distribution provided by \textit{Gaia}, we expect to have a significant number of fast rotators in the sample.
We therefore included in the training set a first sample of 996 targets that includes all \textit{Kepler} stars with periods below 1~day and a random selection of \textit{Kepler} stars with periods spanning from 1 to 15 days. 
We then completed the sample with 996 more stars with no prior selection on the rotation periods, and 995 for which \citet{Santos2019,Santos2021} were not able to perform a measurement of the rotation period. This makes a total of 2987 stars in the training set.

\textit{Kepler} light curves are four-year-long, spanning from 2009 to 2013, while K2 light curves are significantly shorter (about 90~days). In order to ensure that ROOSTER was trained with a set of parameters comparable to the ones we extracted from the K2 light curves, we subdivided our \textit{Kepler} light curves in 90-day long chunks, and decided to consider each of these chunks independently. We then computed the GWPS, ACF, and CS of each of these chunks as outlined in Sect.~\ref{sec:rotation_analysis}.

The training parameters we considered are the following: the candidate rotation periods obtained with each methods, the width of the corresponding Gaussian profiles fitted in the case of the GWPS and CS (which are used as uncertainties on the period), the $S_{\rm ph}$ value computed from each candidate period, its corresponding error, the $H_{\rm ACF}$ and $G_{\rm ACF}$ control parameters \citep[e.g.]{Breton2021}, and the summary parameters of the Gaussian profiles fitted in the GWPS and CS (amplitude, central period, width, uncertainty).

For stars with measured rotation periods, we use a given 90-day chunk in the training set only if one of the candidate rotation periods match the reference values with a 10~\% tolerance, otherwise we remove the chunk from the training sample. ROOSTER is finally trained with 16686 light curve chunks of stars with detected rotation and 10693 light curve chunks of stars without detection. It is tested with 4187 light curve chunks of stars with detected rotation and 2674 light curve chunks of stars without detection.

\section{Comparison with \citet{Reinhold2020} \label{sec:comparison_with_rh}}

\begin{figure}[ht!]
    \centering
    \includegraphics[width=0.49\textwidth]{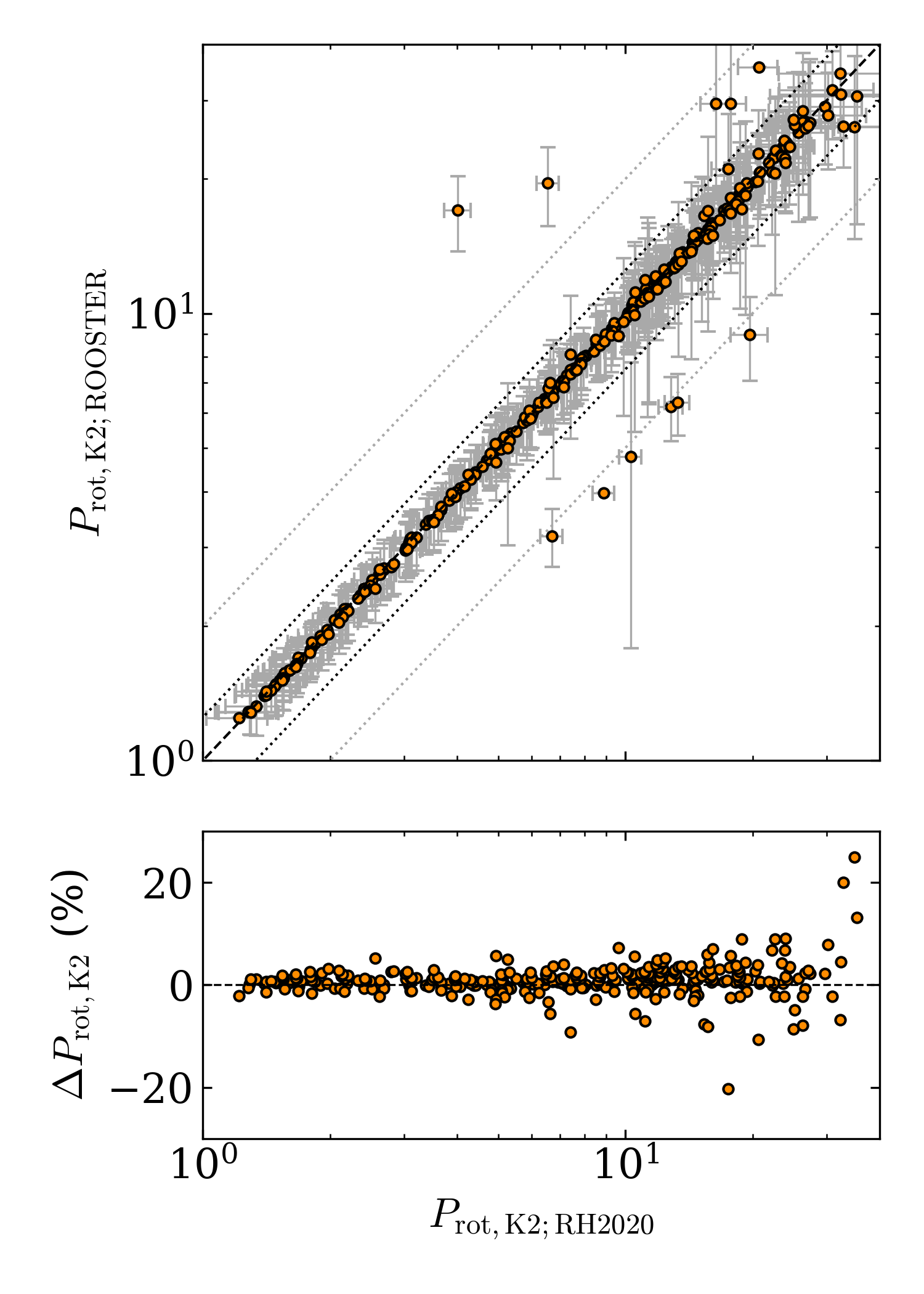}
    \caption{\color{black}Comparison between our measurements, $P_{\rm rot;ROOSTER}$ and the ones from \citet{Reinhold2020}, $P_{\rm rot,K2;RH2020}$. The top panel shows $P_{\rm rot,K2;ROOSTER}$ versus $P_{\rm rot,K2;RH2020}$ while the bottom panel shows $\Delta P_{\rm rot,K2}$ versus $P_{\rm rot,K2;RH2020}$ with $\Delta P_{\rm rot,K2} = (P_{\rm rot,K2;ROOSTER} - P_{\rm rot,K2;RH2020}) / P_{\rm rot,K2;RH2020}$. The 1:0.5, 1:2 (dotted grey lines), 1:0.75, 1:1.25 (dotted black lines) and 1:1 (dashed black line) lines are shown. }
    \label{fig:comparison_reinhold_hekker_2020}
\end{figure}

Some of the stars for which we measured rotation in this work are part of the catalogue published by \citet{Reinhold2020}\footnote{The catalogue is accessible on Vizier: \url{https://cdsarc.u-strasbg.fr/viz-bin/cat/J/A+A/635/A43}.}. Over our 744 stars with measured rotation periods, we found 337 stars in common. As, for each star, \citet{Reinhold2020} provided one value per campaign of observation, when dealing with stars with multiple measurements, we considered only the closest value to our own measurement. 
As it is possible to see it in Fig.~\ref{fig:comparison_reinhold_hekker_2020}, the agreement is very good: out of the 337 stars, 326 measurements are compatible within 25\%. The majority of the stars outside this region (9 stars over 11) are close to the 1:2 and 1:0.5 line, suggesting a possible confusion between the fundamental period and its first overtone.

\section{Summary table}

Table~\ref{table:summary_table} summarises the properties of the 744 targets for which we measured a rotation period in the K2 light curve.

\begin{sidewaystable*}
\caption{Excerpt of the table summarising the properties of the K2 stars studied in this work.}
\label{table:summary_table}
\centering
\begin{tabular}{lcccccccc}
\hline\hline
EPIC & $G$ (mag) & $G_{\rm BP} - G_{\rm RP}$ (mag) & $M_G$ (mag) & $P_\mathrm{rot}$ (day) & Flag & $S_\mathrm{ph}$\tablefootmark{a} (ppm) & $\log R'_{\rm IRT}$ & $\overline {\rho}$ \\
\hline
201094776 & 15.3 & 0.85 & 5.60 & 11.15 $\pm$ 3.00 & 1 & 1831 & - & -0.50 \\
201097220 & 13.6 & 0.82 & 4.59 & 6.08 $\pm$ 1.05 & 0 & 4043 $\pm$ 85 & - & 0.25 \\
201098208 & 14.8 & 1.29 & 6.59 & 13.04 $\pm$ 2.35 & 0 & 5827 & - & 0.20 \\
201102172 & 16.2 & 2.43 & 10.05 & 3.45 $\pm$ -1.00 & 1 & 7464 $\pm$ 932 & - & 0.42 \\
201103864 & 13.4 & 1.06 & 5.89 & 21.74 $\pm$ 4.99 & 1 & 4658 & - & 0.21 \\
201114114 & 14.6 & 1.20 & 6.47 & 16.42 $\pm$ 4.29 & 0 & 1168 & - & 0.04 \\
201116316 & 12.9 & 1.14 & 5.63 & 2.57 $\pm$ 0.45 & 1 & 15747 $\pm$ 7197 & - & 0.87 \\
201121691 & 12.7 & 1.12 & 5.34 & 3.92 $\pm$ -1.00 & 1 & 8178 $\pm$ 2876 & - & 0.82 \\
201123647 & 14.0 & 1.11 & 5.72 & 6.77 $\pm$ 1.10 & 0 & 15041 $\pm$ 2389 & - & 0.70 \\
201131271 & 13.9 & 1.09 & - & 2.25 $\pm$ -1.00 & 1 & 7808 $\pm$ 1771 & - & 0.78 \\
201132458 & 15.0 & 2.24 & - & 2.54 $\pm$ 0.38 & 1 & 14898 $\pm$ 1060 & - & 0.71 \\
201134718 & 13.4 & 1.70 & 7.09 & 1.43 $\pm$ -1.00 & 1 & 8676 $\pm$ 1884 & - & 0.94 \\
201140864 & 16.0 & 2.40 & 10.15 & 5.11 $\pm$ 0.82 & 1 & 19791 $\pm$ 927 & - & 0.33 \\
201149128 & 15.9 & 2.40 & 10.04 & 2.02 $\pm$ -1.00 & 1 & 18889 $\pm$ 587 & - & 0.52 \\
201154929 & 14.8 & 2.07 & 8.69 & 23.59 $\pm$ 4.98 & 1 & 4839 & - & 0.25 \\
201166582 & 15.9 & 1.89 & 8.10 & 1.33 $\pm$ 0.21 & 1 & 2772 $\pm$ 928 & - & 0.30 \\
201178569 & 15.8 & 2.42 & 9.57 & 2.70 $\pm$ 0.42 & 1 & 7211 $\pm$ 1013 & - & -0.10 \\
201212507 & 14.6 & 1.58 & 7.58 & 11.33 $\pm$ 3.80 & 0 & 3994 & - & 0.21 \\
201223827 & 15.8 & 1.16 & 6.60 & 1.86 $\pm$ -1.00 & 1 & 8634 $\pm$ 1354 & - & 0.18 \\
201235266 & 12.5 & 1.22 & 6.44 & 23.20 $\pm$ 5.01 & 1 & 4031 & -5.4 & 0.50 \\
201249315 & 14.9 & 1.13 & 6.17 & 2.70 $\pm$ 0.40 & 1 & 8163 $\pm$ 1781 & - & 0.82 \\
201272989 & 14.3 & 1.58 & 7.48 & 16.16 $\pm$ -1.00 & 1 & 2293 $\pm$ 1264 & - & 0.24 \\
201273096 & 14.9 & 1.33 & 6.45 & 14.95 $\pm$ 4.16 & 0 & 1724 & - & 0.24 \\
201288384 & 10.8 & 1.78 & - & 5.30 $\pm$ 0.82 & 1 & 19583 $\pm$ 1776 & - & 0.80 \\
201308300 & 11.8 & 0.75 & 4.46 & 4.48 $\pm$ 0.83 & 1 & 3121 $\pm$ 720 & -6.1 & 0.67 \\
201346007 & 15.1 & 2.26 & 9.38 & 17.03 $\pm$ 3.27 & 0 & 8271 & - & 0.46 \\
201366994 & 15.0 & 1.35 & 5.73 & 0.54 $\pm$ 0.08 & 1 & 11001 $\pm$ 1141 & - & 0.57 \\
201377225 & 14.4 & 1.90 & 7.61 & 3.18 $\pm$ 0.47 & 1 & 16981 $\pm$ 2943 & - & 0.46 \\
201378489 & 13.7 & 1.52 & 6.63 & 28.11 $\pm$ 12.10 & 0 & 2580 & - & -0.24 \\
201383978 & 17.9 & 2.74 & - & 2.89 $\pm$ 1.13 & 1 & 2161 $\pm$ 625 & - & 0.01 \\
\bottomrule
\end{tabular}
\tablefoot{
A machine-readable version of this table is available. 
The \textit{Flag} column indicates whether the stars belong to the K2-\textit{Gaia} cross validate sample (1) or not (0). 
When possible, the $B_p-R_p$ magnitudes are corrected with the Apsis dereddening coefficients.
\tablefoottext{a}{The $S_{\rm ph}$ activity index being computed on light curve segments of length $\min (t_{\rm obs}, 5 \times P_{\rm rot})$, with $t_\mathrm{obs}$ the total observing time, we do not provide an uncertainty on $S_\mathrm{ph}$ for targets for which the light curve is shorter than $5 \times P_{\rm rot}$.} When possible, the $B_p-R_p$ magnitudes are corrected with the Apsis dereddening coefficients.
}
\end{sidewaystable*}

\end{document}